\pdfoutput=1

\documentclass[10pt,journal,compsoc]{IEEEtran}
\usepackage{nopageno}
\usepackage{color}
\usepackage{listings}
\usepackage{graphicx}
\usepackage{pgffor}
\usepackage{scrextend}
\usepackage{multirow}
\usepackage[table,xcdraw,dvipsnames]{xcolor}
\usepackage{cite}
\usepackage{paralist}
\usepackage{siunitx}
\usepackage{rotating}
\usepackage{bbding}
\usepackage{eqparbox}
\usepackage{pdfpages}
\usepackage{wrapfig}
\usepackage{booktabs}
\usepackage{arydshln}
\usepackage{flushend}
\usepackage{blindtext}
\usepackage{lscape}
\usepackage{hyperref}
\usepackage[multiple]{footmisc}
\usepackage{tikz}
\usepackage{enumitem}
\usepackage{multirow}
\usepackage[linesnumbered,ruled,vlined]{algorithm2e}
\usepackage[framemethod=tikz]{mdframed}
\usepackage{lipsum}
\usetikzlibrary{shadows}
\newmdenv[tikzsetting= {fill=white!20},roundcorner=10pt, shadow=true]{myshadowbox}
\usepackage{graphics}
\usepackage{colortbl}
\usepackage{multirow} 
\usepackage{balance}
\usepackage{picture}
\usepackage[table,xcdraw]{xcolor}
\usepackage{soul}
\usepackage{fourier} 
\usepackage{array,booktabs}
\usepackage{makecell}
\usepackage{times}
\usepackage{wasysym}
\usepackage{xstring}
\usepackage{blindtext}
\usepackage{xcolor}
\usepackage{framed}

\usepackage{verbatim}
\usepackage[export]{adjustbox}
\renewcommand{\footnotesize}{\scriptsize}
\definecolor{lightgray}{gray}{0.8}
\definecolor{darkgray}{gray}{0.6}
\usepackage[table]{xcolor}
\definecolor{Gray}{rgb}{0.88,1,1}
\definecolor{Gray}{gray}{0.85}
\definecolor{Blue}{RGB}{0,29,193}

\newcommand{\sway}{{\sc Sway~}}

\definecolor{co1}{RGB}{249,239,230}
\definecolor{co2}{RGB}{255,192,202}
\definecolor{co3}{RGB}{144,237,144}
\definecolor{MyDarkBlue}{rgb}{0,0.08,0.45} 
\definecolor{Gray}{gray}{0.95}
\definecolor{LightGray}{gray}{0.975}
\newcommand{\bi}{\begin{itemize}}%[leftmargin=0.4cm]}
\newcommand{\ei}{\end{itemize}}
\newcommand{\be}{\begin{enumerate}}
\newcommand{\ee}{\end{enumerate}}
\newcommand{\tion}[1]{\S\ref{sect:#1}}
\newcommand{\fig}[1]{Figure~\ref{fig:#1}}
\newcommand{\tab}[1]{Table ~\ref{tab:#1}}
\newcommand{\eq}[1]{Equation~\ref{eq:#1}}

\newcommand{\resptof}[1]{
}

\newcommand{\respto}[1]{}

\usepackage{url}
\usepackage[framed]{ntheorem}
\usepackage{framed}
\usepackage{tikz}
\usetikzlibrary{shadows}
\theoremclass{Lesson}
\theoremstyle{break}

\tikzstyle{thmbox} = [rectangle, rounded corners, draw=black,
  fill=Gray!40,  drop shadow={fill=black, opacity=1}]

\newshadedtheorem{lesson}{Answer}
\graphicspath{{./imgs/}}

\begin{document}
\title{``Sampling'' as a Baseline Optimizer for Search-based Software Engineering}
\author{Jianfeng Chen,~%~\IEEEmembership{Member,~IEEE,}
        Vivek Nair,~%~\IEEEmembership{Member,~IEEE,}
        Rahul Krishna,~%~\IEEEmembership{Member,~IEEE,}
        and Tim~Menzies,~\IEEEmembership{Member,~IEEE}% <-this % stops a space
        \thanks{The authors
          are with the Department of Computer Science, North Carolina State University, USA.
E-mail: jchen37@ncsu.edu, vivekaxl@gmail.com, rkrish11@ncsu.edu, tim@menzies.us}% <-this % stops an unwanted space
\thanks{Manuscript received  XX, 2017; revised  XX, 2017.}}

\thispagestyle{plain}
\pagestyle{plain}
\IEEEtitleabstractindextext{%
\begin{abstract}
Increasingly, Software Engineering (SE) researchers use search-based optimization techniques to solve SE problems with multiple conflicting objectives.
These techniques often apply CPU-intensive evolutionary algorithms to explore generations of mutations to a population of candidate solutions.  An alternative approach, proposed
in this paper, is to start with a very large population and sample down to
just the better solutions. 
We call this method ``\sway'', short for ``the sampling way''.
\respto{2-1}{{
This paper compares \sway versus state-of-the-art search-based SE tools using  seven models: five software product line models;
and two other software process control models (concerned with   project management, effort estimation, and selection of requirements)
during incremental agile development.
For these models, the experiments of this paper show that \sway is competitive with corresponding state-of-the-art evolutionary algorithms 
while requiring orders of magnitude fewer evaluations.
}}
Considering the simplicity and effectiveness of \sway, we, therefore, propose this approach as
a baseline method for search-based software engineering models, especially for models that are very slow to execute.

\end{abstract}
 
}

% \pagenumbering{arabic} %XXX delete before submission
  
\maketitle 
\IEEEdisplaynontitleabstractindextext
%\IEEEdisplaynontitleabstractindextext
\section{Introduction}\label{sect:introduction}

Software engineers often need to answer questions
that explore trade-offs between competing goals. For example:
\be
\item
What is the smallest set of test cases that cover all program branches?
\item
What is the set of requirements that balances software development cost and customer satisfaction?
% \item What cheapest  resources achieve  most functionality?
\item What sequence of refactoring steps take the least effort while most decreasing the future maintenance costs of a system?
\ee
SBSE, or search-based software engineering, is a commonly-used technique
for solving such problems.
Two things are required for using SBSE methods:
\bi
\item
The {\em   model} is some device which, if its inputs are perturbed,
generates multiple outputs (one for each objective). 
\item
The {\em optimizer} is the
a device that experiments with different model inputs to improve
model outputs. 
\ei
Different models can require different optimizers.
According to Wolpert \& Macready~\cite{Wolpert:1997}, no single algorithm can ever
be best for all optimization problems. They caution that for every
class of problem where algorithm $A$ performs best, there is some other class of problems where $A$  will perform poorly. Hence,
when commissioning a new domain, there is always the need for
some experimentation to match the particulars of the local model
to particular algorithms.

When conducting such commissioning experiments, it is very useful to have a 
{\em baseline} optimizer; i.e., an algorithm which can  generate 
{\em floor performance}  values. Such baselines let a developer
quickly rule out any optimization option that falls ``below the floor''.
 In this way, researchers and industrial practitioners
can achieve fast early results, while also gaining some guidance in all their
subsequent experimentation (specifically: ``try to beat the baseline'').

This paper proposes a new algorithm called \sway (short for the \underline{s}ampling \underline{way}) as a baseline optimizer  for search-based SE problems.
As described in the next section, \sway has all the properties desirable
for a baseline method such as simplicity of implementation and fast execution
times.  Further, the experiments of this paper show that \sway usually performs as well as, or better than, more complex algorithms even for some very hard problems
(e.g., selecting candidate products according to five objectives from highly constrained product lines). Most importantly, \sway adds
very little to the overall effort required to study a new problem.
For example, we tested \sway in three different SE problems: 1) reducing risk, defects as well as development efforts of a project, 2) optimizing agile project structures, and 3) utilizing software product line model to find out features to develop in requirement engineering.
For  all models,  {\sway}'s median cost was just   3\% 
of runtimes  and    1\%  of the number of model
evaluations (compared to only running the standard optimizers).

The rest of this paper is structured as follows:
The remainder of this section will introduce our prior work and main contributions of this paper. Section 2 briefly
introduces the background of SBSE and evolutionary algorithms. Section 3 shows
the core algorithms of {\sc Sway}. In Section~4, \sway is  applied on above three
SE case studies. Results from these case studies are discussed in Section~5. 
After that, the rest of the paper 
 explores threats to validity, reviews some more
related work.

The conclusion of this paper is  {\em not} that \sway is always the best
choice optimizing SBSE models. Rather, since \sway is so simple and so fast, it is a reasonable
first choice for benchmarking other approaches. To aid in that benchmarking process,
all our scripts and sample problems
are available online in Github\footnote{https://github.com/ginfung/fsse}.
Also, to simplify all future references to this material,
the same content has been assigned a digital object identifier
in a public-domain repository\footnote{http://doi.org/10.5281/zenodo.495498}.

\subsection{Relation to Prior Work}
This paper significantly extends prior work of the authors. In 2014,
Krall \& Menzies proposed
GALE~\cite{krall15:hms,galepaper,krall14aaai} that solved
multi-objective problems via a combination of methods.
A subsequent report~\cite{nair16}
found that GALE needlessly over-elaborated some aspects of its design.
That subsequent report evaluated a preliminary version of \sway using results from two similar models, XOMO and POM3 models, both of whose decision representation are a continuous numeric array. That subsequent
the report was expanded into a journal article~\cite{chen2017beyond} to explore a lightly constrained model for Next Release Problem (NRP).
This current paper began when it was realized that the methods used in that journal article failed when applied to {\em heavily constrained models}
and models with {\em binary decisions}.

In {\em heavily constrained models},  a naive ``generate-at-random'' strategy results in too few candidate solutions. Accordingly, this paper processes heavily
constrained models using an SAT solver to generate the initial population.

Our prior versions of \sway used various heuristics to divide the space
of candidates-- all of which fail for models with {\em binary decision variables}. The reason
for this is simple: numeric decisions tend to spread candidates all over the $D$-dimensional space
containing the $D$ decisions. However, for $D$ binary decisions, all the candidates fall to the
vertexes of the $D$-dimensional decision space. Hence, \sway was failing when it kept proposing
useless divisions of the empty space between the vertices. 
Accordingly, to distribute the candidates containing binary decisions, this paper uses
a novel coordinate system. In that coordinate system, initial candidates are first divided by problem-specific heuristics, then grouped by similarities.

Another important distinctive feature of this paper is its evaluation methodology.
In this paper, when evaluating the performance of \sway on our models,
we took care to compare against demonstrably state-of-the-art alternatives.
For example, we do not use the default settings of the NSGA-II~\cite{deb2000fast} optimizer
but instead, apply
an extensive grid search operation to find better settings.

Overall, the unique contributions of this paper are: 

\be

\item A   new baseline approach to multi-objective optimization;
\item Two forms of this new approach: one for continuous variables and
another one for  discrete variables (this discrete version of \sway has not been published before);
\item Results are evaluated by more metrics (Generational Distance, Generated Spread, Pareto Front Size, and Hypervolume);
\item Results are compared against state-of-the-art or highly-tuned algorithms;
  \item Case studies show that \sway 
 allows for a very rapid processing of complex and large heavily constrained models;
 \item Defining an executable method for baselining new SBSE methods. To allow ready access to that method,  our scripts and sample problems are available online for free download. 
 \ee

\section{Background}
\subsection{Baselining with \sway} \label{sect:baseline}
Experienced researchers  endorse the use of  baseline algorithms.
For example, in his textbook on {\em Empirical Methods for AI}, Cohen~\cite{cohen95} strongly advocates comparing supposedly sophisticated systems against simpler alternatives.
In the machine learning community, Hotle~\cite{holte93} uses
the OneR baseline algorithm as a {\em scout}
that runs ahead of a more complicated learner as a way to judge
the complexity of up-coming tasks.
In the software engineering community,  Whigham et al.~\cite{Whigham:2015}
recently proposed baseline methods for effort estimation
(for other baseline methods in effort estimation, see Mittas et al.~\cite{mittas13}). 
Shepperd and Macdonnel~\cite{shepperd12z} argue convincingly that
  measurements are best viewed as ratios compared
  to measurements taken from some minimal baseline
  system.
  Work on cross versus
within-company cost estimation has also recommended
the use of some  very simple  baseline (they recommend
regression as their default model)~\cite{Kitchenham2007}.

In their recent article on baselines in software engineering, Whigham et al.~\cite{Whigham:2015} propose  guidelines for designing a baseline implementation
that include:

\be 
\item Be {\em simple} to describe and implement; 
\item Be {\em applicable to a range of models};
\item Be {\em publicly available} via a reference implementation and associated environment for execution;
\ee
To their criteria, we would add that for multi-objective optimization
algorithms, such baselines should also:
\begin{enumerate}
  \setcounter{enumi}{3}
\item Offer {\em comparable performance} to standard methods. While we do not expect a baseline method to
out-perform all state-of-the-art methods, for a baseline to be insightful, it needs to offer a level of performance
that often approaches the state-of-the-art.
\item {\em Not be resource expensive to apply}
(measured in terms of required CPU or number of evaluations).
The resources required to reach a decision are not a major concern for Whigham's  cost estimation work.
Before a community adopts SBSE baseline methods, we must first ensure that baseline executes very quickly.
Some search-based software engineering methods can require days to years of CPU-time to terminate~\cite{Wang:Harman13}.
Hence, unlike Whigham et al., we take
care {\em not} to select baseline methods that are impractically slow.
\end{enumerate}
\sway satisfies all the above criteria. The method is straightforward:
\bi
\item Generate a very large population of random candidates;
\item Evaluate a small number of representative candidates (using the methods described in \S\ref{sect:sway});
\item Cull any candidates that are near the poorly performing representatives.
\ei  
Note that this uses much less machinery than a standard genetic algorithm;
i.e., there are no complex selection, mutation or crossover operators. Nor
does \sway employ multi-generational reasoning. As a result, it is a simple
matter to code \sway (see pseudocode in Algorithm~\ref{fig:cluster}).

As to being {\em applicable to a wide range of models},
in this paper we apply \sway to models with boolean and continuous decision variables:
\bi
\item
Our models
with continuous decision variables are XOMO and POM3.
 XOMO~\cite{me07f,me09a,me09e} is an  SE model where the optimization task is to reduce the defects, risk, development months, and the total number of staff members associated with a software project.
 POM3~\cite{turner03,port08,1204376} is an SE model of agile development towards negotiating what tasks to do next within a team.
 \item Our model with boolean decision variables is  software product lines~\cite{sayyad13b,sayyad13a}  for which
  the optimization task is to  extract  (a)~valid products that (b) have more features and  use (c)~the most familiar features that (d)~costs the least to implement and which
(e)~has the fewest known bugs.
  \ei
  
% As aside, we note that outside this paper, \sway produced competitive
% results to MOEAs for simulation models of developers working in agile
% projects, the POM3 models~\cite{nair16}. 

As to {\em public availability}, a full implementation of \sway including
all the case studies presented here (including working implementations 
of other multi-objective  evolutionary algorithms   and our evaluation models
in \cite{nair16}) is available online.

In terms of {\em comparative performance}, for each model, we compared \sway's
performance against the established state-of-the-art method as reported in the literature. In those comparative results, 
\sway was usually as good, and sometimes even a little better, than
the state-of-the-art.

\sway is also {\em not resource expensive to apply}.
The algorithm does not evaluate all of its random candidates. Instead, \sway employs a top-down bi-cluster procedure that finds two distant points $X,Y$, then
labels all points according to which of $X,Y$ they are closest to. The
points are then evaluated, and all points near the worst one are culled.
Hence, \sway only evaluates $log(N)$ of $N$ candidates, which makes
it a relatively very fast algorithm:
\bi
\item
Measured in terms of addition model evaluations,
collecting baseline results from \sway would require
\mbox{$\{1,
1,
1,
2,
2,
6,
6,
9,
33\}\%$} 
additional evaluations in various models.
\item
Measured in terms of total runtime, \sway adds
\mbox{$\{1,
1,
1,
1,
1,
1,
2,
2,
3\}\%$}
to the runtime of other optimizers.
\ei
Note that in the above,   values less than 100 denote evaluations that are {\em fewer} and hence {\em better} than other methods. Note also that, usually, \sway terminates very quickly.

This reduction in runtime is an important feature of \sway since some optimization studies can be very CPU intensive.
For example, recent MOEA studies in software
engineering by Fu et al.~\cite{fu2016tuning} and Wang et al.~\cite{Wang:Harman13}
required 22 days and 15 years of CPU time respectively.
While, to some extent, this high CPU cost can be managed via
the use of cloud computing services, those computing environments are extensively monetized so the total financial cost of tuning can be prohibitive. We note that all that money is a wasted
resource if there is a more straightforward way (e.g., \sway) to achieve similar results.

\sway offers many benefits to practitioners and researchers:
\begin{enumerate}[leftmargin=*]
\item Researchers can use results of \sway as the ``sanity checker''. Experiments showed that results of \sway are much better than random configurations and in most times, it is comparable to standard optimizers.
Consequently, when designing new optimizers, researchers can compare their
results to \sway's to see whether their superiority is significant.
\item Practitioners can use \sway to get quick feedback on their adjustments.
For example, in agile development, managers can apply \sway to POM3 to quickly get approximate
changes of developing efforts or costs when they adjust release plans or team personnel, etc.
\end{enumerate}

\subsection{Search Based Software Engineering (\textbf{SBSE})}
\label{sect:sbse}

\sway is our proposed baseline algorithm for search-based software engineering (SBSE). This section reviews the field of SBSE.

Throughout the software engineering life cycle, from requirement
engineering, project planning to software testing, maintenance, and
re-engineering, software engineers need to find a balance between
different goals such as: 
{
\bi
\item  \textit{Software product line optimization}: Sayyad et. al. \cite{sayyad13a} compared several MOEAs, including SPEA2, IBEA, NSGA-II\cite{deb2000fast}, etc, and found that IBEA algorithm performed best in generating valid products
from product line descriptions  (for details see \S \ref{prob:spl}). 
\item \textit{Project planning}:  Ferrucci et al. \cite{ferrucci2013not,sarro2017adaptive} modified the crossover operator in the NSGA-II algorithm and found that their approach (called NSGA-II$_v$) was useful for planning how to make the best use of project overtime.
\item \textit{Test suite minimization}: Wang et al.\cite{wang2013minimizing}  showed that their ``weighted-based'' genetic algorithm significantly outperformed other methods using industrial case study for  Cisco Systems.
\item \textit{Improving defect prediction}: Fu et al.~\cite{fu2016tuning}  and Harman et at.~\cite{harman2014less} report that software quality predictors learned
from data miners can be improved if an evolutionary algorithm first adjusts the tuning parameters of
the learner.
\item \textit{Software clone detectors}:  Wang,  Harman et al.~\cite{Wang:Harman13} report that
the arduous task of configuring complex analysis tools like software clone detectors
can be automated via multi-objective evolutionary algorithms.
\ei
All of these problems can be viewed as  \textit{optimization problems};
i.e., tune the configuration parameters of a model such that, when that model
runs, it generates ``good'' outputs (i.e., output demonstrably
better than other possible outputs). Given the complexities of software engineering,
SE models are often too complicated to prove that output is optimal. For such models,
the best we can do is run multiple optimizers and report the best output seen across
all those optimizers.
}
In the past, due to the simplicity of software structure, developers/experts could make a decision based on their empirical knowledge. For such models of such simple knowledge, it may have been possible
to demand that those outputs are ``optimal''; i.e., there exists no other configuration
such that a better output can be generated.
However, modern software is becoming increasingly complex.
 Finding the optimal solution to such kind of problems may be difficult/impossible. For example, in a project staffing problem, if there
are ten experts available and 10 activities to be accomplished, the
total number of available combinations is 10 billion
($10^{10}$). For such large search spaces,
exhaustively enumerating and assessing all possibilities is impractical.

\begin{figure}
\includegraphics[width=\linewidth]{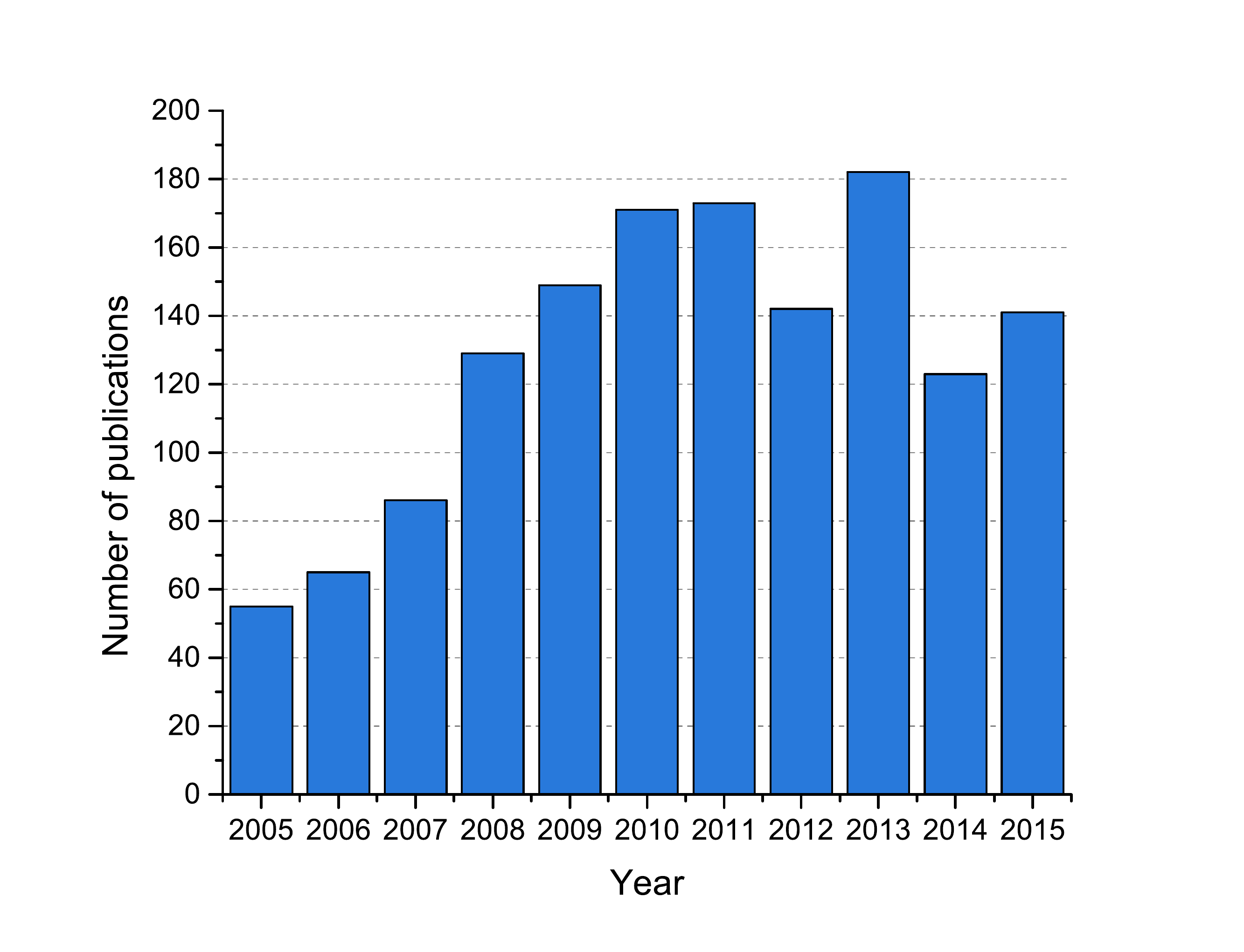}
\caption{SBSE publications. From~\cite{yzmham:sbse-repository}}
\label{fig:sbse_trend}
\end{figure}

When brute force methods fail, it is possible to employ
Metaheuristic search algorithms to explore complex models. The \textit{Search Based
Software Engineering (SBSE)}'s favorite meta-heuristic search algorithms are evolutionary algorithms~\cite{Holland1992}. Such
algorithms offer ``a higher-level procedure or heuristic designed to
find, generate, or select a heuristic (partial search algorithm) that
may provide a sufficiently good solution to an optimization/search problem,
especially with incomplete or imperfect information or limited
computation capacity''~\cite{Bianchi2009}.
As seen in Figure~\ref{fig:sbse_trend}, within the research community, this has become
a very popular approach.

One advantage of these meta-heuristic algorithms is that they can simultaneously explore
multiple goals at the same time. 
The
next section of this paper 
introduces multi-objective evolutionary optimization algorithms, which are widely used in SBSE.

\subsection{Multi-Objective Evolutionary Algorithms (\textbf{MOEA})}
\label{sect:moea}

In SBSE, the software engineering problem is treated as a mathematical
model: given the numeric (or boolean) configurations/decisions
variables, the model should return one or more objectives.  In a
nutshell, the model can convert \textit{decisions} ``\textit{d}'' into
\textit{objective} scores ``\textit{o}'', i.e.
\begin{equation}\label{eq:one}
    \textit{o = model(d)}
\end{equation}
The direction of optimization for the objectives can be to either
maximize or minimize their values.  For example, in software
engineering, we might want to maximize the delivered functionality
while also minimizing the cost to make that delivery.  If model
delivers just one objective, then we call the this a
\textit{single-objective optimization problem}.  On the other hand,
when there are many objectives we call that a \textit{multi-objective
  optimization problem}.

For the multi-objective optimization problem, often there is no 
``$d$'' which can minimize (or maximize) all objectives. Rather, the
``best'' $d$ offers a good trade-off between competing objectives. In
such a space of competing goals, we cannot be optimal on all
objectives, simultaneously.  Rather, we must seek a {\em Pareto frontier} or solution of multiple solutions where no other solutions in the frontier ``dominate'' any other~\cite{zitzler1999multiobjective}.

There are two types of dominance-- \textit{binary dominance} and \textit{continuous dominance}.
{\em Binary dominance} is defined as follows: solution $x$ is said to binary dominate the solution $y$ if and only if the objectives in $x$ is partially less (larger when the corresponding objective is to maximize) than the objectives in $y$, that is,
\begin{center}
\begin{math}
     \forall o  \in \textit{obj }  o_{x} \succeq o_{y}\text{ and } \exists o \in \textit{obj } o_{x} \succ o_{y}
\end{math}
\end{center}
where \textit{obj} are the objectives and ($\succeq,\succ$) tests if an objective score in one individual is (no worse, better) than  the other.  {\em Continuous dominance}~\cite{zitzler2004indicator}, favors $x$ over $y$ if $x$ ``losses'' least:

\begin{equation}\label{eq:cdom}
\begin{array}{rcl}
x \succ y & =& \textit{loss}(y,x) > \textit{loss}(x,y)\\
\textit{loss}(x,y)& = &\sum_j^n -e^{\Delta(j,x,y,n)}/n\\
\Delta(j,x,y,n) & = & w_j(o_{j,x}  - o_{j,y})/n
\end{array}
\end{equation}

MOEAs create the initial population first, and then execute the crossover and mutation repeatedly until ``tired or happy'';
i.e., until we have run out of CPU time limitation or until we have reached solutions that suffice for the purposes at hand.
The basic framework for MOEAs is as follows:
\be
    \item Generate generation 0 using some {\em initialization policy}
    \item Evaluate all individuals in generation 0
    \item Repeat until tired or happy
        \be
        \item {\em Cross-over} items in current generation to make new population;
        \item {\em Mutate} population by making small changes;
        \item {\em Evaluate} individuals in the population;
        \item {\em Select}  some elite subset of the population to form a new generation.
        \ee
        \ee
One simple way to understand MOEAs is to compare them with 
Darwin's theory of evolution. To find good scores for the objectives,
start from a group of individuals. As time goes by, the
individuals inside the group crossover. The offspring which have better fitness scores tend to survive (in the selection step). During the evolution, the mutation operation can increase the diversity of the
group and avoid the evolution from getting trapped in the local optimal.

Standard MOEA algorithms might be not suitable for some SE models. 
% Consequently, researchers reform some operators so that the problems can be solved under the evolutionary algorithm framework. 
For example, the standard \textit{initiation operator} is to build members of the population by selecting, at random, across the range of all known decisions. However, this may not be the best procedure for all models. For example:
\bi
\item
Sayyad et al. found that the best way to seed a population for a five-goal optimization problem
was to first run a  two-goal optimizer (for the hardest pair of goals), then use the two-goal
optimizer as input to the five-goal optimizer~\cite{sayyad13b}. 
\item
Later in this paper (\S\ref{prob:spl}), we will use a case-study of optimizing a heavily constrained model. For that case study, only three out of 10,000 randomly generated candidates satisfied the constraints of that model. Hence, for that model, \sway uses an SAT-solver to initialize the space of candidates. 
\ei
{
There are many  \textit{cross-over} and \textit{mutation} operators, such as one/two point(s) cross-over,
Gaussian mutation, FlipBit mutation,
uniform partially matched crossover (UPMX)~\cite{cicirello2000modeling}, etc.
Specific domains might require specialized cross-over operators.  For example, for program repair, \cite{oliveira2016improved} created a new cross-over operator, {\sc Unif1Space} which improved the fix rate by 34\%.
}

When exploring new MOEAs, much attention has been paid to the {\em selection} operator.
For example, the core innovation in  \mbox{NSGA-II}~\cite{Deb2002} is its method of performing the selection.
Candidates are sorted heuristically into {\em bands} according to how many other candidates they dominate.
The top $B$ bands containing some desired number $N$ candidates survive to the next generation.
If these $B$ bands contain more than $N$ candidates, then:
\bi
\item  NSGA-II sorts   candidates  by  each objective $o_x$.
\item Next,
 NSGA-II annotates each candidate $y$ with the gap $g_o^y$ to its nearest neighbors within the sort
 of objective $o$, where
$o(x) < o_x(y) < o(z)$ and  $g_o^y= o(z) - o(x)$ 
\item  The {\em crowding-distance} $D$ around a candidate $y$ is a hyper-rectangle with volume 
$D_y=\prod_{o\in  \textit{obj }} g_o^y$.
\item NSGA-II sorts the candidates $D_y$  in the last band, then {\em selects} the candidates
 from the {\em least}
crowded regions.\ei
The rationale for this {\em select} rule is as follows.  In crowded regions, we can reject some candidates while
still retaining many others. However, in order to retain the shape of the Pareto frontier, it
is important to retain candidates from the less-crowded regions.

As to the {\em evaluation operator}, the standard approach is, for each decision, run the underlying model to generate objective
scores for those decisions. Such an evaluation operator may be too cumbersome for many reasons:
\bi
\item
Verrappa and Letier warn that ``..for industrial problems, these algorithms generate (many) solutions, which makes the tasks of understanding them  and selecting  one among  them
difficult and time-consuming''~\cite{Veerappa2011}.
\item
Zuluaga et al. comment on the cost of evaluating all decisions for their models of software/hardware co-design:
``synthesis of only one design can take hours or even days.'' ~\cite{Zuluaga:13}.
\item
  Harman comments on the problems  of evolving  a  test suite  for software:  if every
  candidate solution requires a time-consuming execution of the entire system: such test suite generation can take weeks of execution time~\cite{yoo2013gpgpu}.
  \item
    Krall \& Menzies explored the optimization of complex NASA models of air traffic control. After discussing the simulation needs of NASA's research scientists,
    they concluded that those models would take three months to execute, even utilizing NASA's supercomputers~\cite{krall15:hms}.
  \ei
{
Hence, {\sc Sway}'s evaluation operator strives
to reduce the number of requested model evaluation. To achieve this goal,
\sway applies a {\em sampling technique} (discussed in \S\ref{sect:sway}) that reduces the number of model evaluations and, hence,
the total running time.  
}

\section{\sway: The Sampling WAY}
\label{sect:sway}
{
This section introduces our method {\sc Sway} that
recursively clusters the candidates in order to isolate the  superior cluster.
Unlike the common MOEA algorithms (where
 candidates are improved by multiple generations of mutation, crossover, and selection),
 \sway just selects a small superior set candidates among a group of candidates. Consequently, the first step of \sway is to generate a huge amount of candidates. We generated 10,000 in our experiments.

If we cluster the candidates through their {\em objectives}, we need to evaluate all candidates (just like common MOEA algorithms), \sway would be very slow, since model evaluations in many SE problems are extremely time-consuming (see \tion{sbse}). Instead, the \sway clusters the candidates by their {\em decisions}
(recall that  {\em decisions} and {\em objectives} were distinguished in \S\ref{sect:moea} \eq{one}).

Implicit in decision clustering is the assumption that there exists a close association
between the genotype ({\em decision}) and phenotype {\em objective}) spaces. 
In SE, this is not an unwarranted assumption.
 For example,  cloud environment configuration meets such requirement-- an improved number of VM/memories can lead to better quality service   ~\cite{ardagna2014multi}. 
 Also, in the POM3, XOMO model and  software product line model
 (see \S\ref{sect:bmarks}), \sway worked satisfactorily suggesting that at least models have a closely associated genotype/phenotype spaces.
}

\newcommand\mycommfont[1]{\footnotesize\ttfamily\textcolor{black}{#1}}
\SetCommentSty{mycommfont}
\SetKwInOut{Parameter}{Parameter}
\SetKwInOut{Require}{Require Func}
\SetKwInOut{Input}{Input}
\SetKwInOut{Output}{Output}

\begin{algorithm}[!t]
  \scriptsize
  \SetKwFunction{numberof}{numberOf}
\DontPrintSemicolon
\Input{items -- The candidates}
\Output{pruned results}
\Parameter{enough -- The minimum cluster size}
\Require{
{\sc Split}, see \S\ref{sect:split_con}, \S\ref{sect:split_bin}\\
{\sc Better}, see \S\ref{sect:better}}
\BlankLine

\uIf{\numberof{items} < enough}{
    \Return{items}\;
}
\uElse{
    $\Delta_1,\Delta_2 \leftarrow \emptyset,\emptyset$\;
    [west, east], [westItems, eastItems] $\leftarrow$ {\scshape Split}(\textit{items})\;
    \lIf{$\neg${\scshape Better}(west, east)}{
      $\Delta_1  \leftarrow $ \sway(eastItems)}
    \lIf{$\neg${\scshape Better}(east, west)}{
      $\Delta_2  \leftarrow $ {\scshape Sway}(westItems)}
    \Return{$\Delta_1 + \Delta_2$}
}

\caption\sway
\label{fig:cluster}
\end{algorithm}

\noindent
%YYY [TODO assert equal half splitting]

Algorithm \ref{fig:cluster} shows the general framework of {\sc Sway}. The candidates are split into two parts according to their decisions. Then \sway prunes half of them based on the objectives of the corresponding representatives, where the limited number of model evaluations come from.
The {\sc Split} function may differ for different types of decision representation and we will discuss the {\sc Split} function very soon:
\bi
\item
  If the population size is smaller than some threshold, then we just return all candidates (line 1). Otherwise, \sway  splits the
candidates into two parts, ``west side'' and the ``east side.''
\item
After that, lines 6 and 7 compare representatives of the sides. \sway uses different methods to
find those representatives-- see 
\S\ref{sect:split_con} and \ref{sect:split_bin}.
\item
Prune the
candidates based on a comparison of the representatives.
If neither representative is better, then we \sway on each part.
\ei

\sway is a divide-and-conquer process. If the number of candidates is $N$ the number of candidate evaluations is $O(log N)$.

{
The decision spaces in SE models have various types of representations, such as continuous/discrete arrays, graph/tree-based structures, permutations, etc.
The {\sc Split} function is designed according to each of these different types. In this paper, we use two {\sc Split} function, one for continuous decision spaces, another for the binary (and this second split operator is a unique contribution of this paper).
}
\subsection{{\sc Split} for continuous decision spaces}
\label{sect:split_con}
% the SSBSE16 paper method
      {\sc Split} clusters the candidate into parts then picks up representatives for each part. For models with continuous decisions,
      we use the Fastmap heuristic\cite{Platt05fastmap,Faloutsos1995} to quickly split the candidates.
      Platt~\cite{Platt05fastmap} shows that FastMap is a  Nystr\"om algorithm that finds approximations to eigenvectors.

\begin{algorithm}[!t]
  \scriptsize
  \SetKwFunction{sort}{sort}
\DontPrintSemicolon
\Input{items -- The candidates to split}
\Output{[west, east] -- representatives;\\
[westItems, eastItems] -- two parts}
\Require{{\sc Distance}}
\BlankLine

rand $\leftarrow$ randomly selected item in candidates\;
east $\leftarrow$ furthest item apart from rand \tcp{{\sc Distance} required}
west $\leftarrow$ furthest item apart from east \tcp{{\sc Distance} required}
c $\leftarrow$ {\sc Distance}(east, west)\;
\ForEach{x $\in$ \textit{items}}{
a $\leftarrow$ {\sc Distance}(x, west)\;
b $\leftarrow$ {\sc Distance}(x, east)\;
x.d $\leftarrow$ $(a^2+c^2-b^2)/(2c)$ \tcp{cosine rule}
}
\sort items by $x.d$\;
eastItems $\leftarrow$ first half of items\;
westItems $\leftarrow$ second half of items\;
\Return{\text{[west, east], [westItems,  eastItems]}}
\caption{{\sc Split} for continuous space (uses FASTMAP)}
\label{alg:split_con}
\end{algorithm}
\begin{algorithm}[!t]
  \scriptsize
  \SetKwFunction{counts}{count}
\SetKwFunction{sort}{sort}
\SetKwFunction{max}{max}
\SetKwInOut{Parameter}{Parameter}
\DontPrintSemicolon
\Input{items -- The candidates to split}
\Output{[west, east] -- representatives;\\
[westItems, eastItems] -- two parts}
\Parameter{totalGroup -- the granularity}
\Require{{\sc Distance}}
\BlankLine

rand $\leftarrow$ randomly selected item in candidates\;
\ForEach{x $\in$ \textit{items}}{
x.r $\leftarrow$ $| \forall d_i \in x \wedge x==1 |$ \tcp{sum all the ``1'' values}
x.d $\leftarrow$ {\sc Distance}(x, rand)\;
}
normalize $x.r$ into $[0,1]$\;
R $\leftarrow$ $\{$i.r | i $\in$ \text{items}$\}$ \tcp{all possible radius}
\ForEach{k $\in$ R}{
\tcp{for each possible radius}
\tcp{equally distribute the candidates with k-radius into the concentric-circle}
g $\leftarrow$ $\{i|i.r = k\}$\;
\sort g by $x.d$. g $\leftarrow$ $x_1,x_2,\ldots, x_{|g|}$\;
\For{i $\in$ $[1,|g|]$}{
$x_i.\theta$ $\leftarrow$ $\frac{2\pi i}{|g|}$\;
}
}
\tcp{split the candidates}
thk $\leftarrow$ \max(R)/(totalGroup) \tcp{the thickness}
\ForEach{a $\le$ totalGroup}{
\tcp{for the annulus with $(a-1)$thk$\le radius \le a $ thk}
g $\leftarrow$ $\{i|(a-1) $thk$ \le i.r \le a $thk$ \}$\;
$c_1$ $\leftarrow$ the item with minimum $\theta$ in g\;
add $c_1$ to east\;
$c_2$ $\leftarrow$ the item with maximum $\theta$ in g\;
add $c_2$ to west\;
add items with $\theta \le \pi$ in $g$ to eastItems\;
add items with $\theta > \pi$ in $g$ to westItems\;
}
\Return{\text{[west, east],[westItems, eastItems]}}
\caption{{\sc Split} for binary decision spaces}
\label{alg:split_bin}
\end{algorithm}
%YYY check line number (checked --jf) 
Algorithm \ref{alg:split_con} lists the details of {\sc Split} function. To split
the candidates into two parts according to the FastMap heuristic,
first, pick any random candidate (line 1) and then find the two extreme candidates based on the distances (line 2-3). The {\sc Distance} used in our case studies is the \textit{Euclidean distance}. All other candidates
are then projected onto the line joining the two extreme candidates(line 5-8). Finally,
split the candidates into two parts based on their projection in the line.

\subsection{{\sc Split} for binary decision spaces}
\label{sect:split_bin}

\sway using Algorithm \ref{alg:split_con}  performed well on models with numerical decisions. 
However, when applied to the problem with binary decisions, i.e., $D=\{0,1\}^n$, it was observed that
\sway performed far worse than standard MOEAs. 
On investigations, we found that for $D$ binary decisions, all the candidates fall to the
vertices of the $D$-dimensional decision space. Hence, the continuous version of Split described in the last section
was failing when \sway kept proposing
useless divisions of the empty space between the vertices.

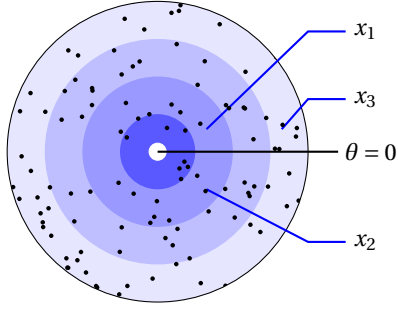
\begin{figure}
\centering
\begin{tikzpicture}
\pgfmathsetseed{24782015}
\draw[fill=blue!10] (2,2) circle (2 cm);
\fill[fill=blue!25] (2,2) circle (1.5 cm);
\fill[fill=blue!40] (2,2) circle (1 cm);
\fill[fill=blue!65] (2,2) circle (0.5 cm);
\draw[white,fill=white] (2,2) circle (0.75ex);
\draw[black, thick] (2,2) -- (4.4, 2);
\node[text width=1cm] at (5, 2) {$\theta=0$};

\draw[blue, thick] (2.65, 2.3) -- (4, 3.6) -- (4.5, 3.6);
\draw[blue, thick] (2.65, 1.5) -- (4, 0.8) -- (4.5, 0.8);
\draw[blue, thick] (3.65, 2.3) -- (4, 2.7) -- (4.5, 2.7);
\node[text width=1cm] at (5.1, 3.6) {$x_1$};
\node[text width=1cm] at (5.1, 0.8) {$x_2$};
\node[text width=1cm] at (5.1, 2.7) {$x_3$};
\begin{scope}
\clip (2,2) circle (2 cm);
\foreach \i in {1,...,600}{
\fill [fill=black](rand*4,rand*4) circle (0.03);
}
\end{scope}
\end{tikzpicture}

\caption{Map candidates into a circle. The large white dot in the
  center is the ``pivot'', selected randomly among the candidates. 
  The horizontal black line denotes $\omega=0$.
  All
  other candidates (the black dots) are located based on their radius and angular coordinates. The circle is divided into multiple equal-thickness annuli. The candidates with minimum angular coordinates form the \textit{east} representatives. The candidates with maximum angular coordinates form the \textit{west} representatives. Candidates whose angular
  coordinate is less than $\pi$(upper semicircle) form the
  \textit{eastItems} and others (locates in lower semicircle) form the
  \textit{westItems}.
 }
\label{fig:splitting2}
\end{figure}

Accordingly, inspired by the research in radial basis function kernel\cite{chung2003radius}, we applied a radial coordinate system.
This co-ordinate for vectors of binary decisions forces them away from outer edges into the inner volume
of the decision space. Candidates representing similar-size individuals (i.e., that share a similar number of  ``1'' bits) are grouped, and comparisons only are performed inside the group.
Algorithm \ref{alg:split_bin} implements such a  radial co-ordinate system.
  This algorithm splits our binary decision using a randomly
selected ``pivot'' point.  After that, it maps the other candidates
into a \textit{circle}, rather than the \textit{line} showed in
Algorithm \ref{alg:split_con}.

To map the candidates into this circle, first, for the candidate
$x=(d_0,d_1,\ldots,d_n)$ ($d_i\in\{0, 1\}$), we assign $x.r$ as
$\sum_{i=0}^n d_i$ and $x.d$ as the \textit{Jaccard distance} between
$x$ and the ``pivot'' candidate(lines 2-4). The Jaccard distance between A and B is
defined as
\[\sum  |a_i-b_i| \quad 0\le i \le n  \]
where $A=(a_0,\ldots,a_n), B=(B_0,\ldots,B_n)$ and $ a_i,b_i \in \{0,1\}$.

This Jaccard distance is a common distance measurement for binary strings.
Similar to Euclidean distance applied in \S\ref{sect:split_con}, the Jaccard distance
is easy to compute and satisfies the triangle inequality~\cite{kosub2016note} -- one of the requirements for metric space.

Once mapped into a
circle, we then uniformly spread all candidates with similar $r$ values
into a
circumference whose radius is $r$, based on their $d$ values-- the
one with minimum $d$ values has the minimum angular coordinate; the
one with second minimum $d$ values has a larger coordinate, and so on
(lines 7-11).
For example:
\bi
\item
Suppose this split procedure randomly  selects $r=(1,0,1,1,0)$  as the pivot.
Using this pivot, we can place 
  \mbox{$x_1=(0,0,1,1,0)$}, \mbox{$x_2=(0,1,0,0,1)$} and \mbox{$x_3=(1,1,1,1,0)$} onto Figure \ref{fig:splitting2} as follows. 
\item
$x_1,x_2,x_3$ contain $\{2,2,4\}$ ``1'' values (respectively).
  Hence, these are placed in rings 2 and 4 of Figure \ref{fig:splitting2}. 
\item $x_2,x_3$ are the most similar, different (respectively) to the pivot $r$. Hence, these vectors
get the smallest, largest $\theta$ values from the black line in Figure \ref{fig:splitting2} that denotes $\theta=0$.
\ei  
This circle is then used to generate the  partitions.
Figure \ref{fig:splitting2} shows how this circle is divided into several equal-thickness
annuli (the number of annuli, i.e., the granularity of {\sc Split}
is a configurable parameter). After the division:
\bi
\item
  The candidates with minimum
  $\theta$ in each annulus area form the \textit{east};
\item
  The
  candidates with maximum $\theta$ form the \textit{west}.
  \item
Candidates in
the upper semicircle form the \textit{eastItems} and others form the
\textit{westItems}.
\ei

\subsection{The {\sc Better} function and Other Parameters}
\label{sect:better}
The {\sc better} function is for comparing the representatives for two halves of the candidates. \sway uses binary domination for individual comparisons. When the representatives are paired into different groups (such as in Algorithm \ref{alg:split_bin}), and if there are more pairs such that east representative dominates west representatives, then \sway prunes the west half (and \textit{vice versa}).

The {\em enough} parameter in Algorithm \ref{fig:cluster} controls the size of final cluster. \sway set it as $\sqrt{N}$, where $N$ is number of total candidates, i.e. 10,000 in our experiments.

The ``totalGroup'' is the granularity in Algorithm \ref{alg:split_bin}. 
Some engineering judgment is required to set this parameter. For this paper,
we tried 2, 4, 6,..., 20 and found no improvement in Hypervolume\footnote{As described in  \S\ref{sect:measures}, Hypervolume is diversity  as well as convergence indicator used to assess MOEAs.} after a value of 10. Hence, that value was used for the rest of this paper.

\subsection{Application of \sway  Other SBSE Problems}\label{sect:extension}
\respto{2-2}{Based on our experience with  {\sc Sway}, 
this section  offers guidelines on how to apply this algorithm in 
different domains. }

In the following, we will say a model:
\bi
\item
is {\em numeric} if its decision
variables range across the number plans; e.g. ``age'' would be numeric.
\item
is {\em discrete} if the decision
variables come from a small range;
e.g. ``days of week'' would be discrete.
\item
is {\em boolean} if the decision variables are discrete {\em and}
have the range {\em true, false}.
\item
is {\em highly constrained} if more than 50\% of randomly generated
solutions do not satisfy the constraints of that model.
\ei
Using that terminology, we can offer the following guidelines on how to
use {\sc Sway}. 
\bi
\item As described in \cite{nair2016accidental}, use Algorithm~\ref{alg:split_con} for {\em numeric} models.
\item
As described in \cite{chen2017beyond}, for {\em discrete} models, 
first, find a way to do a coarse-grain grouping of the decisions
(e.g., for decision that follows some temporal sequence, group the earlier decisions before the later decisions). Once decisions are so grouped,
apply  Algorithm~\ref{alg:split_con} to each grouping.
\item As described in this paper, for {\em boolean, highly constrained} models,
use a radial coordinate system for the decisions and an SAT solver to
generate the initial population sample.
\ei

Another frequently asked question relates to our use of SAT solver technology. The whole point of \sway is that it is a simple baseline method.
If such a   method requires an SAT solver, does it not negate
the ``simplicity'' requirement of a baseline method?

To answer this second question, we note that SAT is a very
mature technology.   This work used   PicoSAT solver, which is a python package that can be readily installed using the standard package mangers. Once installed, it took less than an hour to make that code accept the CNF formulae, and then to return candidate items for
our initial population. 

\section{Case Studies}
\label{sect:case_studies}
{
To explore and analyze the efficiency of {\sc Sway}, we compared \sway with commonly used MOEA algorithms under three benchmarks. In this section, we will briefly introduce these three benchmarks, including the model definition and related research work for each of them,
and then the exploration process to several research questions. Finally, we describe the performance measures we used in our experiments.
}

\subsection{Benchmarks}\label{sect:bmarks}
% This section reviews the three SE problems studied in this
% paper--XOMO, POM3 and software product lines (SPL). They differs in the type of decision representation as well as lies in various software development stages.
% , and lies in effort estimation, project management and require engineering respectively.

\respto{2-3}{
In selecting case studies for this paper, we reflected over the space of model types seen in the SBSE literature.
The following is an approximate characterization of those models:
\bi
\item Model size: large or small;
\item Conflicting constraints: many or few;
\item Decision types: discrete or continuous.
\ei
Our reading of the literature is that:
\be
\item
The most frequently seen are {\em small} models with {\em continuous} decisions and 
\underline{no} constraints.
\item
The hardest models to solve, that are most used to stress test MOEAs, are  {\em large discrete} models
with {\em many conflicting constraints}.
\ee
For our evaluation,
we selected  models
that fall across the
range of the above model types.
For example:
\bi
\item
The software product lines discussed below have discrete-valued decisions and many conflict constraints.
\item
At the other extreme, the XOMO model discussed below is a much smaller model with continuous-valued decisions and no
constraints.
\item
In between these two extremes, we added the  POM3 model (that used continuous-valued decisions) since prior work showed
that POM3 is very slow to optimize~\cite{galepaper}.
\ei
Another reason to use the models described below is the existence of prior results from   these models~\cite{krall15:hms,galepaper,krall14aaai,lekkalapudi2014cross,sayyad13b}.
% , published at senior SE forums. 
That is, by using the particular models described below, we can compare \sway to the state-of-the-art.

Note that some consideration was given to using artificially generated models that could better span the space of models size, constraints,
and decision types. In prior work, we used such models~\cite{menzies2002applications, owen2002makes, owen2002alternative, menzies1997applications, menzies2000test}, but based on feedback from the SE community,
we can no longer endorse that approach. Specifically, the space of possible artificially generated models is so huge that it is
difficult to show that results from any artificial model are relevant to any specific model.
}

\begin{figure}[!t]
{\scriptsize
  \renewcommand{\baselinestretch}{0.9}
\begin{center}
\begin{tabular}{r|r@{:~}l}\toprule
scale factors  &prec & have we done this before?\\
 (exponentially  &flex & development flexibility \\
   decrease effort)     &resl & any risk resolution activities?\\
       &team &  team cohesion\\
        &pmat & process maturity\\\hline
upper  &acap & analyst capability\\
(linearly decrease       &pcap & programmer capability\\
 effort)     &pcon & programmer continuity\\
       &aexp &  analyst experience\\
      &pexp &  programmer experience\\
      &ltex &  language and tool experience\\
      &tool &  tool use\\
      &site &  multiple site development\\
      &sced & length of schedule   \\\hline
lower &rely &    required reliability  \\
(linearly increase     &data &   2nd memory   requirements\\
 effort)     &cplx &  program complexity\\
      &ruse &  software reuse\\
      &docu &   documentation requirements\\
      &time &   runtime pressure\\
      &stor &   main memory requirements\\
     &pvol &    platform volatility  \\
\end{tabular}
\end{center}} 
\caption{Descriptions of the XOMO variables.}
\label{fig:emsf2}
\end{figure}
\begin{figure}[!t]
  \renewcommand{\baselinestretch}{0.9}
{\scriptsize
\begin{center}
\begin{tabular}{l|lrr|lr}
      &\multicolumn{3}{c|}{ranges}      &\multicolumn{2}{c}{values}\\\hline
project&feature&low&high&feature&setting\\\hline
 &rely&3&5&tool&2\\
FLIGHT:&data&2&3&sced&3\\
 &cplx&3&6&&\\
JPL's flight&time&3&4&&\\
software&stor&3&4&&\\
&acap&3&5&&\\
&apex&2&5&&\\
&pcap&3&5&&\\
&plex&1&4&&\\
&ltex&1&4&&\\
&pmat&2&3&&\\
&KSLOC&7&418&&\\ 
\multicolumn{6}{c}{~} \\\hline
 &rely&1&4&tool&2\\
GROUND:&data&2&3&sced&3\\
 &cplx&1&4&&\\
JPL's ground&time&3&4&&\\
software&stor&3&4&&\\
 &acap&3&5&&\\
&apex&2&5&&\\
&pcap&3&5&&\\
&plex&1&4&&\\
&ltex&1&4&&\\
&pmat&2&3&&\\
&KSLOC&11&392&&\\
\multicolumn{6}{c}{~} \\\hline
&prec&1&2&data&3\\
OSP:&flex&2&5&pvol&2\\
 &resl&1&3&rely&5\\
Orbital space &team&2&3&pcap&3\\
plane nav\&&pmat&1&4&plex&3\\
gudiance&stor&3&5&site&3\\
  &ruse&2&4&&\\
 &docu&2&4&&\\
&acap&2&3&&\\
&pcon&2&3&&\\
&apex&2&3&&\\
&ltex&2&4&&\\
&tool&2&3&&\\
&sced&1&3&&\\
&cplx&5&6\\
&KSLOC&75&125\\
\multicolumn{6}{c}{~} \\\hline
 &prec&3&5&flex&3\\
OSP2: &pmat&4&5&resl&4\\
 &docu&3&4&team&3\\
OSP&ltex&2&5&time&3\\
version 2&sced&2&4&stor&3\\
 &KSLOC&75&125&data&4\\
 &&&&pvol&3\\
 &&&&ruse&4\\
 &&&&rely&5\\
&&&&acap&4\\
&&&&pcap&3\\
&&&&pcon&3\\
&&&&apex&4\\
&&&&plex&4\\
&&&&tool&5\\
&&&&cplx&4\\
&&&&site&6
\end{tabular}
\end{center}}
\caption{Four project-specific XOMO case studies.}\label{fig:xomocases}
\end{figure}

\subsubsection{XOMO}

XOMO, introduced in \cite{menzies2005xomo}, is a general framework for
Monte Carlo simulations that combine four COCOMO-like software
process models from Boehm's group at the University of Southern
California.  Figure \ref{fig:emsf2} lists the description of XOMO
input variables (All should be within $[1,6]$). The XOMO user begins
by defining a set of \textit{ranges} or a specific \textit{value} of
these variables to address his or her real situation in one software
project. For example, if the project has (a)~{\em relaxed schedule
  pressure}, they should set {\em sced} to its minimal value; (b)~{\em
  reduced functionalists}, they should halve the value of {\em kloc}
and minimize the size of the project database (by setting {\em
  data=2}); (c)~{\em reduced quality} (for racing something to
market), they might move to lowest reliability, minimize the
documentation work and the complexity of the code being written,
reduce the schedule pressure to some middle value-- in the language of
XOMO, this last change would be {\em rely=1, docu=1, time=3, cplx=1}.

XOMO computes four objective scores: 
(1)~project {\em risk}; 
(2)~development {\em effort}; (3)~predicted {\em defects}; (4)~total {\em months} of development.
Effort and defects are predicted from mathematical models derived from data collected from hundreds
of commercial and  Defense Department projects~\cite{boehm00b}. 
 As to the {\em risk} model, this model contains
 rules that trigger when management decisions decrease the 
odds of completing a project: e.g.,  demanding
{\em more}  reliability ({\em rely}) while  {\em decreasing} analyst capability ({\em acap}).
Such a project is ``risky'' since it means the manager is demanding more reliability from less skilled analysts.
XOMO measures {\em risk} as the percent of triggered rules.

The optimization goals for XOMO are to:
\be
\item Reduce risk;  
\item Reduce effort;
\item Reduce defects;
\item Reduce months.
\ee
Note that this is a non-trivial problem since the objectives listed above as non-separable and conflicting. For example, {\em increasing} software reliability {\em reduces} the number of added defects while {\em increasing} the software development effort. Also,
{\em more} documentation can improve team communication and {\em decrease} the number of introduced defects. However, such increased
documentation {\em increases} the development effort.
{Consequently, XOMO is multi-objective optimization problem. MOEA algorithms can handle this. \cite{galepaper} and \cite{lekkalapudi2014cross} pointed out that the NSGA-II~\cite{deb2000fast} can solve this problem and return quite good results. In our experiments, we will compare \sway with the NSGA-II algorithm in solving XOMO cases.
}

In our case studies with XOMO, we use four scenarios taken from NASA's Jet
Propulsion Laboratory~\cite{me09a}. As shown in \fig{xomocases},
FLIGHT and GROUND are general descriptions of all JPL flight and
ground software while OSP and OPS2 are two versions of the flight
guidance system of the Orbital Space Plane. 

From \fig{xomocases}, we can know that the FLIGHT model is more flexible than other cases, that is, the decision space for FLIGHT is larger than the GROUND or OSPs. Similarly, sorting the decision space of the cases, we have
\begin{equation}\label{eqn:xomo_cmpr}
\text{OSP} \approx \text{OSP2} < \text{GROUND} < \text{FLIGHT}
\end{equation}

\subsubsection{POM3-- A Model of Agile Development}
\label{sect:pom3}

\begin{figure}
  \scriptsize
  
  \renewcommand{\baselinestretch}{0.9}
    \begin{tabular}{|l|l|p{1.3in}|}
        \hline
        Short name &Decision             & Description                                          \\ \hline
        Cult&Culture              & Number (\%) of requirements that change. \\\hline
        Crit&Criticality           & Requirements cost effect for safety critical systems.\\\hline
        Crit.Mod&Criticality Modifier & Number of (\%) teams affected by criticality.           \\ \hline
        Init. Kn&Initial Known        & Number of (\%) initially known requirements.               \\ \hline
        Inter-D&Inter-Dependency     & Number of (\%) requirements that have interdependencies to other teams.             \\\hline
        Dyna&Dynamism             & Rate of how often new requirements are made.           \\ \hline
        Size&Size            & Number of base requirements in the project.\\        \hline
        Plan&Plan                 & Prioritization Strategy:
        0=~Cost Ascending;  1=~Cost Descending; 2=~Value Ascending; 3=~Value Descending;
        4=~~$\frac{Cost}{Value}$ Ascending.
%Note that a standard agile strategies use ``Value Descending'', i.e. plan=3~\cite{me09j}.
\\\hline
     T.Size&Team Size            & Number of personnel in each team                       \\ 
        \hline
    \end{tabular}
    \caption {List of inputs to POM3.
    These inputs come from Turner \& Boehm's   analysis of factors
    that control how well organizers can react to agile development practices~\cite{turner03}.}
    \label{fig:pom3decisions}
\end{figure}

\begin{figure}%[!t]
  \scriptsize
  \renewcommand{\baselinestretch}{0.9}
 
\begin{center}
    \begin{tabular}{r|p{0.65in}|p{0.65in}|p{0.65in}}
                     & POM3a                         & POM3b             &POM3c       \\ 
                             & A broad space of projects. & Highly critical small projects& Highly dynamic large projects\\\hline
        Culture              & [0.10, 0.90]       & [0.10, 0.90]  & [0.50, 0.90]  \\ 
        Criticality          & [0.82, 1.26]       & [0.82, 1.26]   & [0.82, 1.26]  \\ 
        Criticality Modifier & [0.02, 0.10]       & [0.80, 0.95] & [0.02, 0.08]   \\ 
        Initial Known        & [0.40, 0.70]       & [0.40, 0.70]  & [0.20, 0.50]  \\ 
        Inter-Dependency     & [0.0, 1.0]       & [0.0, 1.0]  & [0, 50] \\ 
        Dynamism             & [1, 50]      & [1.0, 50.0]  & [40, 50] \\ 
        Size & [30, 100]  & [3, 30]     &[30, 300]   \\ 
        Team Size            & [1, 44]      & [1, 44]  & [20, 44]    \\ 
        Plan                 & [0, 4]             & [0, 4]    & [0, 4]      
\end{tabular}
\end{center}

\caption{Three specific POM3 scenarios. }\label{fig:POM3abcd}
\end{figure}

POM3 model is a tool for exploring the  management challenge of agile development\cite{turner03, port08, 1204376}-- balancing   {\em idle rates}, {\em completion rates} and {\em overall cost}. More specifically,
\bi
\item In the agile world, projects terminate after achieving a {\em completion rate} of $X$\% $(X<100)$ of its required tasks;
\item Team members  become {\em idle} if forced to wait for a yet-to-be-finished task from other teams;
\item To {\em lower}  the {\em idle rate} and {\em improve} the {\em completion rate}, management can hire staff--but this {\em increase} the  {\em overall cost}.
\ei 
The POM3 model simulates the Boehm and Turner model of agile programming \cite{boehm00b} where teams select tasks as they appear in the scrum backlog. Figure \ref{fig:pom3decisions} lists the inputs of
POM3 model. What users feel interested in is how to tune the decisions to:
\bi
\item increase completion rates,
\item reduce idle rates,
\item reduce overall cost.
\ei 

One way to understand POM3 is to consider a set intra-dependent requirements.
 A single requirement consists of a prioritization {\em value} and a {\em cost}, along with a list of child-requirements and dependencies.  Before any requirement can be
    satisfied, its children and dependencies must first be satisfied.
    POM3 builds a requirements heap with prioritization values,
    containing
     30 to 500 requirements, with costs from 1 to 100 (values 
     chosen in consultation with Richard Turner~\cite{turner03}). 
Since POM3 models agile projects, the   {\em cost, value} figures are constantly changing (up until the point when the requirement
is completed, after which they become fixed). 
 Now imagine a mountain of requirements hiding below the surface of a lake; i.e., it
is mostly invisible. As the project progresses, requirements (and their dependencies) becomes visible
to the on-looking

Programmers are organized into teams. 
Every so often, the teams pause to plan their next sprint.
At that time, the backlog of tasks comprises   the visible requirements. 
For their next sprint, 
teams prioritize work for their next sprint using one of five prioritization methods:  
(1)~cost ascending;  (2)~cost descending; (3)~value ascending; (4)~value descending; (5)~$\frac{\textit{cost}}{\textit{value}}$ ascending.
Note that prioritization might be sub-optimal due to the
changing nature of the requirements {\em cost, value} as the unknown
nature of the remaining requirements.
POM3 has another wild card; it contains an {\em early cancellation probability} that can cancel a project after $N$ sprints~(the
value directly proportional to number of sprints). Due
to this wild-card, POM3's teams are always racing to deliver as
much as possible before being re-tasked.  The final total cost is a function
    of:
    \bi
    \item[(a)] Hours worked, taken from the {\em cost} of the requirements; \item[(b)] The salary of the developers: less
    experienced developers get paid less;
    \item[(c)] The criticality of the software: mission-critical software costs
    more since they are allocated more resources for software quality tasks.
  \ei

In our study, we explore three scenarios proposed by Boehm personnel
communication (\fig{POM3abcd}). Among them, POM3a covers a wide range
of projects; POM3b represents small and highly critical projects and
POM3c represent large projects that are highly dynamic, where cost and
value can be altered over a large range. 
{
According to Lekkalapudi's report~\cite{lekkalapudi2014cross}, the POM3c is the most complex model among them, or
\begin{equation}\label{eqn:pom3_cpmr}
\text{POM3a} < \text{POM3b} < \text{POM3c}
\end{equation}
}

Similar to the XOMO benchmark, this is also a multi-objective optimization problem. From Lekkalapudi's report, NSGA-II can solve this problem and return quite good results. Consequently, same as the XOMO series, we will compare \sway with the NSGA-II algorithm in solving POM3 study cases.

\begin{figure}[!t]
 \includegraphics[width=\linewidth]{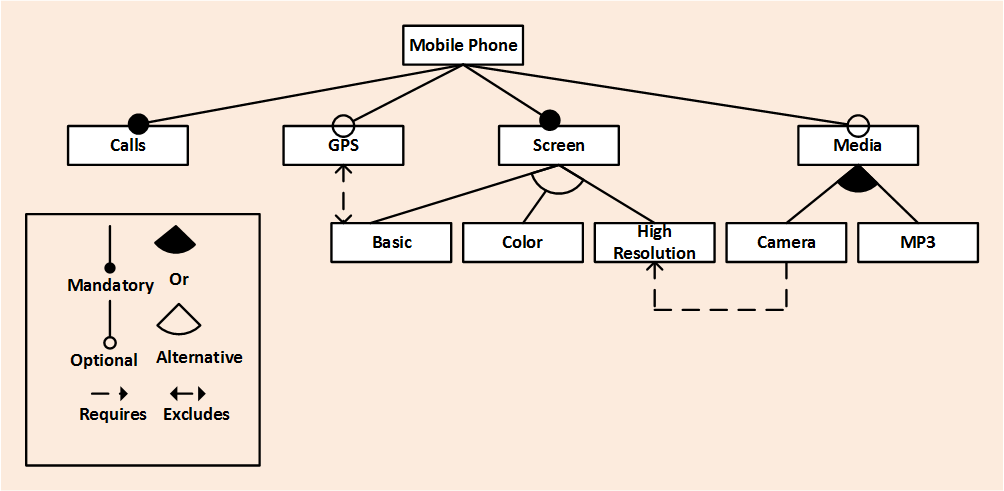}
\caption{Feature model for mobile phone product line. To form a mobile phone, ``Calls'' and ``Screen'' are the mandatory features(shown as \textit{solid $\bullet$}), while the ``GPS'' and ``Media'' features are optimal(shown as \textit{hollow $\circ$}). The ``Screen'' feature can be ``Basic``, ``Color'' or ``High resolution'' (the \textit{alternative} relationship). The ``Media'' feature contains ``camera'', ``MP3'', or both (the \textit{Or} relationship).}
\label{fig:mobile}
\end{figure}

\begin{figure}
{\scriptsize
  \begin{eqnarray*}
    \left\{
    \begin{array}{ll}
        \neg \text{Mobile Phone} \lor \text{Calls}\\
        \text{Mobile Phone} \lor \neg \text{Calls}\\
        \neg \text{Mobile Phone} \lor \text{Screen}\\
        \text{Mobile Phone} \lor \neg \text{Screen}\\
        \text{Mobile Phone} \lor \neg \text{GPS}\\
        \text{Mobile Phone} \lor \neg \text{Media}\\
        \text{Media} \lor \neg \text{Camera}\\
        \text{Media} \lor \neg \text{MP3}\\
        \neg \text{Media} \lor \text{Camera} \lor \text{MP3}\\
        \text{Screen} \lor \neg \text{Basic}\\
        \text{Screen} \lor \neg \text{Color}\\
        \text{Screen} \lor \neg \text{High resolution}\\
        \neg \text{Screen} \lor \text{Basic} \lor \text{Color} \lor \text{High resolution}\\
        \neg \text{Basic} \lor \neg \text{Color} \lor \neg \text{High resolution}\\
        \text{Basic} \lor \neg \text{Color} \lor \neg \text{High resolution}\\
        \neg \text{Basic} \lor \text{Color} \lor \neg \text{High resolution}\\
        \neg \text{Basic} \lor \neg \text{Color} \lor \text{High resolution}\\
        \neg \text{GPS}\lor \neg \text{Basic}\\
        \neg \text{Camera}\lor \text{High resolution}
    \end{array}
    \right.
\end{eqnarray*}}
\caption{\protect\fig{mobile} expressed as CNF}\label{fig:cnf}
\end{figure}

\subsubsection{Software Product Lines}
\label{prob:spl}
A software product line (SPL) is a collection of related software
products, which share some core
functionality~\cite{harman2014search}. From one product line, many products
can be generated.  For example,
Apel et al.~\cite{siegmund2012predicting} model the compilation configuration parameters of databases
as a product line. By adjusting those configurations, a suite of different database
solutions can be generated.

Figure \ref{fig:mobile} shows a feature model for a mobile phone
product line. All features are organized as a tree. The relationship
between two features might be ``mandatory'', ``optional'',
``alternative'', or ``or''. Also, there exist some cross-tree constraints, which means the preferred features are not in the same sub-tree. These cross-tree constraints complicate the process of exploring feature models\footnote{Without cross-tree constraints, one can generate products in linear time using a top-down traversal of the feature model.}. In practice, all constraints, including the tree-structure constraints and the cross-tree constraints can be expressed by the CNF
(conjunctive normal form). For example, Figure \ref{fig:mobile}
can be expressed as the set of CNF formulas shown in  \fig{cnf}.
%YYY not ``figure'' but ``Figure'' (checked by jf)

{
Researchers who explore these kinds of models~\cite{sayyad13a, sayyad13b, harman2014search, henard2015combining}
define a ``good'' product as the one that satisfies
five objectives:

\be
\item Find the valid products (products not violating any cross-tree constraint or tree structure) which have.
\item More features; and
\item Less known defects; and
\item Less total cost; and 
\item Most features used in prior applications.
\ee
 Following the approach of Sayyad~\cite{sayyad13b}, defect, cost, and knowledge of usage in prior applications
is set stochastically. 
}

A major problem with analyzing software product lines
is that it can be very hard to find valid product since real-world software product lines can be
far more complex than our example.
Some software product line models comprise up to tens of thousands
of features,  with 100,000s of constraints (see \tab{spl_scenarios}). These
networks of constraints can get so complex that random assignments of ``use'' or ``do not use'' to the features
have a very low probability of satisfying the constraints. For example, in one of our software product
lines, the \textit{linux} model, we generated 10,000 random sets of decisions for the features. Within that space, less than four decisions were valid.
%YYY which model (checked)

Consequently, 
much of the research on optimizing the generation of products from a software product
lines have focused on how best to optimize within these heavily-constrained models:
\bi
\item
  Sayyad et al.~\cite{sayyad13b}  introduced
  the \textit{IBEASEED} method-- a five-goal optimization problem
  had its first generation of candidates initialized by a pre-processor
  that just sought out valid products (and one other goal).
\item  \textit{SATIBEA} was introduced by Henard et
al. \cite{henard2015combining}. This makes full use of SAT solver technology
to fix the invalid candidates every time the ``mutate'' or
``crossover'' operation is performed in the IBEA algorithm. Results
showed that the SATIBEA algorithm could find the valid products for the
extremely large feature models by tens of thousands evaluations, much
better than other algorithms.
\ei

{To the best of our knowledge, the SATIBEA method is the best algorithm to find the optimal SPL which are represented in the form of CNFs. 
Consequently, we will compare \sway with SATIBEA algorithm.}

{We used five product line models from SPLOT and LAVT repositories. \tab{spl_scenarios} shows the basic information of our five cases. According to the number of features or constraints, the size of decision spaces of five cases are subjected to
\begin{equation}\label{eqn:spl_cpmr}
\text{webportal} < \text{eshop} < \text{fiasco} \approx \text{freebsd} < \text{linux}
\end{equation}
}

\newcolumntype{L}{>{\centering\arraybackslash}m{1.1cm}}
\newcolumntype{M}{>{\centering\arraybackslash}m{2cm}}
\begin{table}
\scriptsize
\caption{Feature models used in this study, sorted by the number of constraints. The constraints include the tree-structure and cross-tree constraints. SPLOT models can be found at \url{http://www.splot-research.org/} and LVAT models are at \url{https://code.google.com/archive/p/linux-variability-analysis-tools/}}
\centering
\rowcolors{2}{white}{gray!15}
\begin{tabular}{MLLc}
    \toprule
    Name(Source) & Number of features & Number of constraints & Reference\\
    \midrule
    webportal (SPLOT) & 49 & 81 & \cite{mendoncca2008decision}\\
    eshop (SPLOT) & 330 & 506 & ~\cite{lau2006domain}\\
    fiasco (LVAT) & 1638 & 5,228 & ~\cite{berger2012variability}\\
    freebsd (LVAT) & 1396 & 62,138 & ~\cite{she2011reverse}\\
    linux (LVAT) & 6888 & 343,944 & ~\cite{she2011reverse}\\
    \bottomrule
  \end{tabular}
\label{tab:spl_scenarios}
\end{table}

\begin{table}
\scriptsize
\caption{Parameters tuned by grid search for the NSGA-II algorithm in solving XOMO and POM3 cases.}
\centering
\rowcolors{2}{white}{gray!15}
\begin{tabular}{MLLL}
    \toprule
    Name & MU & CXPB & MUTPB\\
    \midrule
    OSP & 200 & 0.9 & 0.1 \\
    OSP2 & 100 & 0.8 & 0.2 \\
    GROUND & 200 & 0.8 & 0.15 \\
    FLIGHT & 300 & 0.9 & 0.15 \\
    POM3a & 300 & 0.8 & 0.15 \\
    POM3b & 160 & 0.9 & 0.1 \\
    POM3c & 200 & 0.9 & 0.2 \\
    \bottomrule
  \end{tabular}
\label{tab:nsga2_par}
\end{table}

\subsection{Research Questions}
\label{sect:rq}
{
 To explore {\sc Sway}, we organized our exploration around the following research questions (RQ):
}
\begin{enumerate}[leftmargin=0cm,label={},itemindent=0cm,labelwidth=\itemindent,align=left]
\item \textbf{(RQ1)} To what extent is \sway faster than  typical MOEAs?
\item \textbf{(RQ2)} Can \sway find the solution with maximized (minimized) objective as other MOEAs?
\end{enumerate}

RQ1 questions how fast  \sway is while RQ2 questions whether the results from \sway are comparable to other MOEA algorithms.   Equation (\ref{eqn:xomo_cmpr}), (\ref{eqn:pom3_cpmr}) and (\ref{eqn:spl_cpmr}) indicates the order of 
problem size. In the following, all results will be presented in that size order.

\newcolumntype{Q}{>{\centering\arraybackslash}m{0.82cm}}
\newcolumntype{R}{>{\centering\arraybackslash}m{0.5cm}}
\newcolumntype{T}{>{\centering\arraybackslash}m{1.5cm}}
\begin{table}[!t]
\scriptsize
\caption{Parameter configurations overview}
\centering
\begin{tabular}{Q|Q|R|Q|Q|R|T}
    \hline
    Strategy & Algorithm & Pop Size & Crossover Rate & Mutation Rate & Repeat & Termination\\\hline
    \multirow{2}{*}{MOEA} & NSGA-II & \multicolumn{3}{c|}{See Table~\ref{tab:nsga2_par}} & 30 & Not improved in 5 gens\\ \cline{2-7}
    & SATIBEA & 300$\dagger$ & 0.05 & 0.001* & 30& Not improved in 5 gens\\\hline
    Sample & SWAY & 10,000 & / & / & 30 & See Algorithm~\ref{fig:cluster} \\\hline
    Brute-Force & Groud-Truth & 10,000 & / &/&30 & Generation 0\\\hline
    
  \end{tabular}
  \begin{flushright}
  $\dagger$ archive size = pop size = 300\\
  * Standard mutate: 0.001; Smart mutate: 0.98\\
  POM3 and XOMO models are optimized by NSGA-II, SWAY and GroundTruth\\
  SPL models are optimized by SATIBEA, SWAY and GroundTruth
  \end{flushright}
\label{tab:all_para}
\end{table}

{
There are many MOEA algorithms or modified MOEA to solve our benchmarks. 
In the following, we compare   \sway against some arguably state-of-the-art methods.
Our reading of the literature is that:
\bi
\item
Best prior results for XOMO and POM3 were reported using  NSGA-II~\cite{deb2000fast,galepaper};
\item
Best prior results for configuring products from product lines were obtained using SATIBEA~\cite{henard2015combining}.
\ei}

When applying  SATIBEA to the software product line models, we used
the control parameters suggested by  Henard et al. \cite{henard2015combining}.
As to applying NSGA-II to XOMO or POM3,
prior reports \cite{galepaper} and \cite{lekkalapudi2014cross} did not state
their control parameters. To adddress that issue:
\bi
\item
We ran a  grid search~\cite{bergstra2012random} for
the three parameters: \textit{population size (MU)}, \textit{cross-over probability (CXPB)} and \textit{mutation probability (MUTPB)}. 
\item
Our grid search space was defined as 
$\{\text{MU}=[100, 120,\ldots, 300]\}\times\{\text{CXPB}=[0.9,0.8,\ldots,0.6]\}\times\{\text{MUTPB}=[0.1,0.15,\ldots,0.25]\}$.
\item We use hypervolume (see \S\ref{sect:measures}) as the quality indicator when grid searching. Here we used hypervolume since it is
the combination of convergence and diversity indicator.
\item
\tab{nsga2_par} shows the tuned parameters. 
\ei

Parameter tuning is necessary for SE exploration, especially in
the area of search-based SE. For example, in a very recent FSE'17 paper, Fu et al.~\cite{fu2017easy} found that naive learner, e.g., SVM, with parameter tuning can even outperform
complex deep learning techniques.

However, when discussing this work with other researchers and colleagues, 
one common question is ``why grid search?''.
Our answer is that: ``grid search'' is simple enough.
Even though researchers created many parameter tuners, for example, Fu~\cite{fu2017easy} applied differential evolutionary optimizer, Arcuri~\cite{arcuri2013parameter} applied the central composite design to
reduce the number of explored configurations, etc., 
grid search can cover most of the possible configurations.

In the following,
whenever we mention NSGA-II, this  will be that algorithm plus the parameters of
\tab{nsga2_par}.

All of our experiments were implemented using the \textit{DEAP}~\cite{DEAPJMLR2012} 
MOEA Python framework. In SATIBEA and the candidate initialization of SPL candidates, an SAT solver is required. Henard et.al \cite{henard2015combining} used the SAT4j solver. 
{
In this paper, we used PicoSAT~\cite{NiemetzPreinerBiere-SMT-Competition-2016} instead. 
We used PicoSAT since it recently achieved impressive comparative performance 
results in an international   \textit{SAT-Race 2015} competition~\cite{balyo2016sat}.
}

{\resptof{3a}
The termination of \sway is controlled by the minimum cluster size (see \tion{better}).}
The termination of  NSGA-II and SATIBEA can be defined as maximum running time, a number of evolution generations or even manual termination, etc. 
{
In this paper, to enable a fair comparison with the state-of-the-art algorithm, we set the termination of existing MOEA algorithm as the point where solution
quality does improve for consecutive five generations.
Here, quality was measured by combing convergence and diversity using the hypervolume metric (see \tion{measures}).
}

{\resptof{1c}
Finally, since there is no mutation or cross-over operations in \sway, all results are from initial random-generated candidate sets. To address the tradeoff of \sway, we also used the NSGA-II selection operator to find the Pareto frontier among the set of initial candidates. This calculation was time-consuming since it needs to evaluate all candidates (10,000 in our experiments) and sorted them.  
In the following, we call this method
the   {\sc GroundTruth}. 
Table~\ref{tab:all_para} concludes all parameters.
}

\subsection{Performance  Measures}
\label{sect:measures}
Our research questions concern how fast the \sway runs and how good the results are.
To explore how fast of \sway (efficiency), we record following two measures.

\begin{enumerate}[leftmargin=0cm,label={},itemindent=0cm,labelwidth=\itemindent,align=left]
\item {\textbf{(M1) Runtime:}} 
The execution time from one algorithm starts to the terminal of that algorithm. Running time is a direct method to compare different method.

\item \textbf{(M2) Number of evaluations:} 
Sometimes comparing the running time is not enough. While all of our methods were coded in the same language (Python), some of the implementation is more mature (and have been optimized better) than others. Since, part of runtimes, we also
record the number of evaluations.
\end{enumerate}

~\\
{\resptof{3c}}
To measure how good the result of \sway are (effectiveness), we followed the guidance of a recent
 ICSE'16~\cite{wang2016practical} paper. That guide advises to record
 four quality measures: generational distance, generated spread, pareto front size and hypervolume. Here we first define $PF_c$ and $PF_0$. $PF_c$ is the Pareto front obtained by an algorithm while $PF_0$ is the optimal Pareto Front for a specific problem. In SE models, it is unfeasible to
obtain the optimal  Pareto Front ~\cite{Deb2001}. Hence, we collected all solutions found by any algorithm and picked up all non-dominated solutions to form the $PF_0$. This strategy is widely applied in the area of MOEA applications~\cite{wang2016practical}.
~\\
\begin{enumerate}[leftmargin=0cm,label={},itemindent=0cm,labelwidth=\itemindent,align=left]
\item {\textbf{(M3) Generational Distance (GD):}} 
GD is a measure for {\em convergence}.
It is the Euclidean distance between solutions in $PF_c$ and nearest solutions in $PF_0$~\cite{Veldhuizen98evolutionarycomputation}. It can be calculated by
\begin{equation}
GD=\frac{\sqrt{\sum^{|PF_c|}_{i=1}d(x_i, PF_0)^2}}{|PF_c|}
\end{equation}
where $d(x_i, PF_0)$ refers to the minimum Euclidean distance from solution $x_i$ in $PF_c$ to $PF_0$. A lower GD indicates the result is closer to the Pareto frontier of a specific problem. A value of 0 means that all obtained solutions are optimal;
i.e. {\em lower} values of GD are {\em better}.

\item {\textbf{(M4) Generated Spread (GS):}} 
GS~\cite{zhou2006combining} is a {\em diversity} indicator. It is defined to extend the quality indicator \textit{Spread} which only works for bi-objective problems. GS can be calculated by
\begin{equation}
GS = \frac{\sum^{m}_{i=1}d(e_i,PF_c)+\sum_{x\in PF_c}|d(x,PF_c)-\bar{d}|}{\sum^{m}_{i=1}d(e_i,PF_c)+|PF_c|*\bar{d}}
\end{equation}
where $(e_1, e_2,\ldots,e_m)$ refers to $m$ extreme solutions for each objective in $PF_0$; $d(*,\dagger)$ refers to the minimum Euclidean distance from solution $*$ to the set $\dagger$; $\bar{d}$ is the mean value of $d(x,PF_c)$ for all solutions in $PF_c$. A lower value of GS shows that the results have a better distribution;
i.e. {\em lower} values  of GS are {\em better}.

\begin{figure*}
\centering
\includegraphics[width=\textwidth]{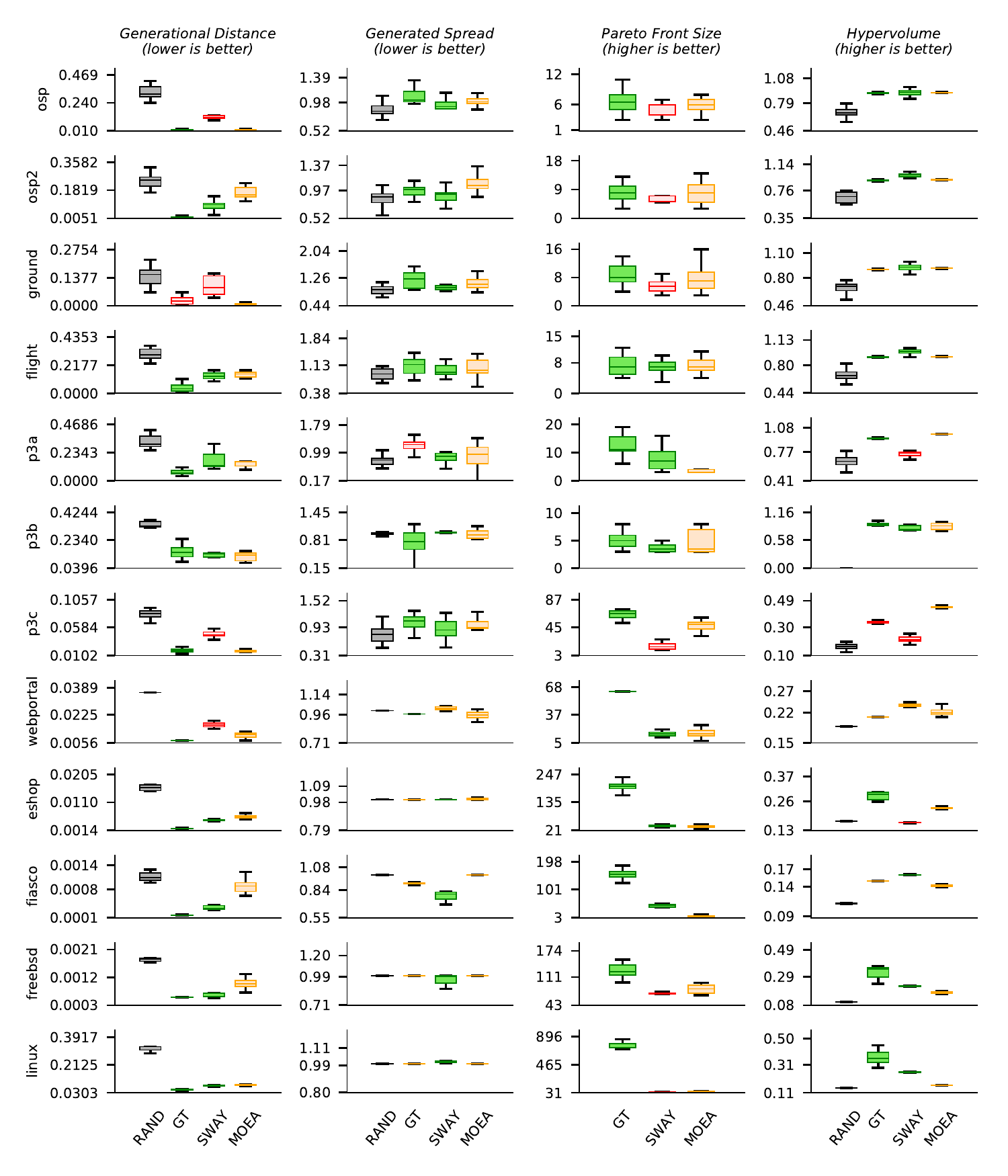}
\caption{The box-plot of four quality indicators in each study cases (30 runs).
{In a boxplot, the middle ``box'' represents the middle 50\% of results among all 30 runs and the line dividing ``box'' into two parts shows the median of all runs. The upper/lower ``whiskers'' mark the extreme results outside middle 50\% of all runs.}
In each plot: \textbf{RAND}=``sanity check'' -- randomly generated candidates (size = MOEA's pareto front size); \textbf{GT}= {\sc GroundTrue}   (the frontier of 10k random generated populations); 
\textbf{\sway}= the method proposed in this paper; \textbf{MOEA}=the prior state-of-the-art MOEA  for
that corresponding study case.
Recall that ``generational distance'' is a coverage indicator;  ``generated spread'' and ``Pareto front size'' are  diversity indicators;
and ``hypervolume'' is a combination of convergence and diversity. 
Note that {\em higher} values are {\em better} for ``Pareto front size'' and ``hypervolume''
while {\em lower} values are {\em better} for ``generational distance'' and ``generated spread''.
Using a Wilcoxon test (5\% significance level) we color these results as follows:
 {\color{Bittersweet}\bf ORANGE} boxes  mark results that are  statistically insignificantly  different from state-of-the-art method; and {\color{ForestGreen}\bf GREEN},    {\color{red} \bf RED } marks  results that are statistically significant better, worse (respectively)  than the state-of-the-art.  For a summary of these results, see \protect\fig{sum1} and \protect\fig{sum2}.
}
\label{fig:boxplot}
\end{figure*}

\item {\textbf{(M5) Pareto front size (PFS):}} 
PFS is another {\em diversity} indicator. It measures the number of solutions  included in $PF_c$, i.e. $|PF_c|$. A higher PFS means that the users
have more options to choose, that is, a more diverse obtained Parto front;
i.e. {\em higher} values of PFS are {\em better}.

\item \textbf{(M6) Hypervolume (HV):}
HV is the {\em combination} of convergence and diversity indicator. 
As defined in \cite{zitzler1999multiobjective}, HV measures the size of space the obtained frontier covered. Formally,
\begin{equation}
\label{equ:hv}
\text{Hypervolume}=\lambda\left(\bigcup_{x\in PF_c}\{x'|x\prec x'\prec x_{\text{ref}}\}\right)
\end{equation}
where
  $\lambda(\cdot)$ is the Lebesgue measure, the standard way measures the subset of n-dimensional Euclidean space. For example, the Lebesgue measure is the length, area or volume when the number of dimensions is  $n=1,2,3$ respectively;
 $\prec$ is the binary domination comparator;
 $x_{\text{ref}}$ denotes a reference point which should be dominated by all obtained solutions.
  Note that, in this study, all objectives are normalized to $[0,1]$ and set $x_{\text{ref}}=(1,1,\ldots,1)$. Notice that a higher value of HV demonstrates a better performance of $PF_c$;
 {\em higher} values of Hypervolume are {\em better}.

\end{enumerate}

{\resptof{3d}\resptof{3g}
To test the performance robustness and reduce the observational error,
we repeated these case studies {30 times} with 30 different random seeds. To simulate real practice, such random seeds are used in initial population creation as well as the successive process.
To check the statistical significance of the differences between the algorithms, we performed a statistical test using Wilcoxon test at a 5\% significance level. Wilcoxon test is a non-parametric test and suitable for the samples when the population cannot be assumed to be normally distributed.
}

\newcommand{\yy}[1]{\cellcolor{co3} {\textit{#1}}}

\newcommand{\nn}[1]{\cellcolor{co2} {\textit{#1}}}

\newcommand{\sAM}[1]{\cellcolor{co1} {\textit{#1}}}

\newcolumntype{B}{>{\centering\arraybackslash}m{0.25cm}}

\begin{table}[!b]
\caption{Median value of runtime and model evaluation numbers from 30 independent runs.
Numbers rounded to the nearest percent (so ``0\%''
means ``less than 0.005'')
}
\centering
\begin{tabular}{r|ccc|ccc}
\cline{2-7}
~  & \multicolumn{3}{c|}{\bf Runtime(seconds)} & \multicolumn{3}{c}{\bf \# Evaluations}\\
 &  \begin{tabular}{B}\sway\\($R_S$)\end{tabular} &\begin{tabular}{B}MOEA\\($R_M$)\end{tabular} & $\frac{R_S}{ R_S+R_M}$ 
 & \begin{tabular}{B}\sway\\($E_S$)\end{tabular} &\begin{tabular}{B}MOEA\\($E_M$)\end{tabular} & $\frac{E_S}{E_S+E_M}$ \\
\midrule
osp & 3.23 & 320 & 1\% & 18 & 4630 & 0\% \\
osp2 & 3.31 & 112 & 3\% & 16 & 2432 & 1\%\\
ground & 3.29 & 388 & 1\% & 20 & 5321 & 0\%\\
flight & 5.62 & 663 & 1\% & 33 & 10980 & 1\%\\\hline
POM3a & 4.69 & 450 & 1\% & 26 & 3836 & 1\%\\
POM3b & 5.01 & 583 & 1\% & 32 & 4532 & 1\%\\
POM3b & 6.33 & 990 & 1\% & 40 & 8302 & 0\%\\\hline
webportal & 6 & 244  & 2\% & 134 & 5100 & 3\%\\
eshop & 18 &321  & 5\% & 136 & 6732 & 2\% \\
fiasco & 63&1065 & 6\% & 122 & 20102 & 1\% \\
freebsd & 188 & 2058 & 8\% & 136 & 26146 & 1\%\\
linux & 1103 & 3295 & 25\% & 168 & 8900 & 2\%\\
\bottomrule

\end{tabular}

\label{tab:runtime}
\end{table}
\begin{figure}[!b]
\begin{center}
{\scriptsize
\begin{tabular}{rr|c|c|c|c}
&          & Generational & Generated & Pareto  & Hyper-   \\
n&model     & Distance     & Spread    & Front Size   & volume    \\\hline
1&osp       &    \yy{0.5}          &        \yy{1.95}   &    \yy{0.46}    &     \yy{0.6}       \\
2&ops2      &    \yy{1.30}         &     \yy{0.82}      &   \yy{0.72}     &    \yy{0.65}        \\
3&ground    &     \nn{0.79}        &      \yy{0.86}     &    \yy{0.71}    &   \sAM{0.30}         \\
4&flight    &    \yy{1.96}         &   \yy{0.86}        &     \yy{0.42}     &   \yy{0.52}         \\
5&pom3a     &     \yy{0.79}         &   \nn{0.53}         &      \yy{01.30}  &    \yy{0.85}        \\
6&pom3b     &        \yy{0.75}     &    \yy{0.43}        &   \yy{0.56}      &  \yy{0.85}           \\
7&pom3c     &     \yy{0.69}         &   \yy{0.42}         &     \yy{0.82}    &  \nn{1.03}          \\
8&webportal &    \yy{0.63}          & \yy{0.90}          &    \yy{1.86}    &    \sAM{0.63}         \\
9&eshop     &    \yy{0.74}         &     \sAM{0.29}      &    \yy{1.23}    &     \yy{1.29}        \\
10&fiasco    &      \yy{1.97}       &        \sAM{0.49}    &    \yy{1.99}    &        \sAM{0.23}    \\
11&freebsd   &     \yy{0.79}        &      \sAM{0.22}      &    \yy{1.98}    &     \yy{0.63}        \\
12&linux     &     \yy{0.72}        &     \sAM{0.13}      &  \yy{1.99}      &        \yy{1.97}    \\\hline
&same + better& 11/12           & 11/12         & 12/12     & 11/12
\end{tabular}}
\end{center}
\caption{{\sc GroundTruth} vs state-of-the-art: How often is ground truth worse, same, or better?
Summarized from 
\fig{boxplot}. Color patterns are the same as \fig{boxplot}. Decimal in each cell is the effect size.  Generating and evaluating all the models in this figure took 52 hours of CPU.}\label{fig:sum1}
\end{figure}
\begin{figure}[!b]
\begin{center}
{\scriptsize
\begin{tabular}{rr|c|c|c|c}
&          & Generational & Generated & Pareto  & Hyper-   \\
n&model     & Distance     & Spread    & Front Size   & volume    \\\hline
1&osp       &    \nn{1.97}           &        \yy{0.45}   &    \nn{0.36}    &     \yy{0.63}       \\
2&ops2      &    \yy{0.99}         &     \yy{0.56}      &   \nn{0.46}     &    \yy{0.49}         \\
3&ground    &     \nn{0.93}         &      \yy{0.60}     &    \nn{0.46}     &   \yy{0.64}         \\
4&flight    &    \yy{0.72}         &   \yy{0.63}        &     \yy{0.49}     &   \yy{0.59}          \\
5&pom3a     &     \yy{0.67}         &   \yy{0.50}         &      \yy{1.93}   &    \nn{1.58}        \\
6&pom3b     &        \yy{0.67}     &    \yy{0.84}        &   \yy{0.69}     &  \yy{0.66}           \\
7&pom3c     &     \nn{1.57}         &   \yy{0.54}        &     \nn{1.44}    &  \nn{1.41}           \\
8&webportal &    \nn{1.97}          & \sAM{0.63}           &    \yy{0.69}    &    \sAM{1.00}         \\
9&eshop     &    \yy{0.78}          &     \yy{0.67}       &    \yy{0.49}    &     \nn{1.56}        \\
10&fiasco    &      \yy{1.23}        &        \yy{1.63}    &    \yy{0.96}    &        \yy{0.99}    \\
11&freebsd   &     \yy{0.78}        &      \yy{0.64}      &    \nn{0.88}    &     \yy{1.53}        \\
12&linux     &     \yy{1.04}        &     \yy{0.69}      &  \nn{0.63}      &        \yy{1.92}    \\\hline
&same + better& 8/12           & 11/12         & 6/12     & 8/12
\end{tabular}}
\end{center}
\caption{\sway vs state-of-the-art: How often is \sway worse, same, or better? Summarized from 
\fig{boxplot}.Color patterns are the same as \fig{boxplot}. Decimal in each cell is the effect size.  Generating and evaluating all the models in this figure took 
the runtimes seen in Table~\ref{tab:runtime}.}\label{fig:sum2}
\end{figure}

\section{Results}
\label{sect:results}

\subsection{RQ1: is \sway Faster than Typical MOEAs?}\label{sect:arq1}
We compared the speed of the algorithms through their runtime as well as the number of model evaluations.
\tab{runtime} shows the median runtime and evaluation numbers recorded in our experiments.
As can be seen in all cases, \sway is faster than the state-of-the-art evaluation algorithms,
often by two orders of magnitude (especially for the smaller models at the top of the table).
And even for the largest models, \sway offered some speed up advantages.
For example, in the study case \textit{linux}, the median runtime of \sway was 1103s (18min), while the runtime of SATIBEA algorithm was near an hour.

Consequently, from \tab{runtime}, our answer to  RQ1 is \sway usually terminates orders of magnitude faster of the other algorithms used in this evaluation.

\subsection{RQ2: Are \sway's Solutions as good as Other Optimizers?}\label{sect:arq2}
\resptof{3e}\fig{boxplot} shows boxplot of quality indicators among the 30 independent runs in our experiment (where
each run used a different random number seed). In that figure:
\bi
\item The RAND method is a ``sanity check'' for our technology. In this approach, 
$N$ candidates were generated at random and then pruned to a final frontier by discarding any dominated points
(domination computed from \tion{moea}). Note that,  to select $N$ for this method, we  used the strategy of~\cite{ouni2017search}:
i.e., set $N$ to the median size of the frontier generated by was used by MOEA.
\item
The {\sc GroundTruth} method, introduced in \tion{rq}, generates and evaluates many solutions,  then reports
the best parts of the Pareto frontier (found using the NSGA-II sorting strategy).
\item 
The \sway method randomly generates solutions, and then fetch some of them through the sampling strategies
described in \S\ref{sect:sway}. Note that \sway only evaluates a very small number of solutions.
\item
The {\sc MOEA} method refers to the state-of-the-art method defined for each case study.
Recall from \S\ref{sect:bmarks} that this state-of-the-art is one of  (a)~the grid-search-tuned NSGA-II or (b)~combining the SAT solver
and IBEA algorithm.
\ei
In \fig{boxplot}, the colors denote a statistical comparison with the state of the art,
where the statistical test is a non-parametric Wilcoxon comparisons at a 5\% significance level:
\bi
\item
{\color{ForestGreen}\bf GREEN}, {\color{red}\bf RED} denote results at are better, worse  (respectively) than the state-of-the-art; 
\item
Results that statistically insiginficantly different to the state-of-the-art are marked in 
{\color{Bittersweet}\bf ORANGE}.
\ei
\fig{sum1} and \fig{sum2} summarize \fig{boxplot} using the same color scheme.
As shown in \fig{sum1},  {\sc GroundTruth} often found best results.
At first glance, the results of
\fig{sum1} seem to say that all work on heuristic multi-objective optimization
should halt since merely making up lots of random solutions performs comparatively
very well indeed. However, not shown in \fig{sum1} is the CPU time to achieve
those results. Recall from Table~\ref{tab:runtime} that \sway required just under 25 minutes to optimize all the models of this paper (and 80\% of that time
was spent on the largest {\em linux} model). By way of comparison,
evaluating all the solutions in \fig{sum1} required 52 hours; i.e., that approach was 124 times slower.  \fig{sum2} comments on the value of the solutions
achieved via that very slow random {\sc GroundTruth} method vs {\sc Sway}.

\fig{sum2} summarizes the results achieved by {\sc Sway}.
 In the majority case,
 across all performance measures, \sway performs the same or better as the state-of-the-art.
 Note that these results were achieved
with the number of evaluations seem in
 \tab{runtime}; i.e. after merely dozens to a few hundred evaluations.  

{One quirk in \fig{boxplot} is that sometimes, very simple {\it RAND} method achieved comparable generated spread values to MOEA or \sway. This is due to the nature of solutions in multi-dimensional space. As noted by Domingos,   random points in large dimensions space are often very distant~\cite{domingos2012few}. Hence, it is not surprising that a random selection does very well (as measured by spread).
Note that achieving good spread scores via random methods says nothing about the {\em value} of the optimizations achieved via that method (merely the dispersion of those candidates). For a comment on the value of the optimization achieved, see the generational distance and hypervolume results of Figure~\ref{fig:boxplot} where, as we would expect,    RAND   performs much worse than other optimizers}.

\section{Threats to Validity}
\label{sect:threats}
\subsection{Optimizer bias}\label{sect:optbias}

The goal of this paper was not to prove that \sway is the best optimizer for all models.
Rather, our goal was to say that, compared to current practice in the literature, \sway offers competitive solutions
at a small fraction of the evaluation costs of other methods.
 Hence, we propose \sway as a  reasonable first choice for benchmarking other approaches.
 
 For that goal, it is not necessary to compare \sway against all other optimizers. Rather, \sway should
 be compared against known state-of-the-art in the literature.

\subsection{Internal bias}
Internal bias originates from the stochastic nature of multi-objective optimization algorithms. The evolutionary process required many random operations, same as the \sway introduced in this paper. 

To mitigate these threats, we repeated our experiments for 30 runs and reported the median/boxplot of the indicators. We also employed statical tests to check the significance of the achieved results.

\subsection{Sampling bias}

This paper studied the performance of \sway using three classes of models:  XOMO, POM3, and
software product lines.  There are many other
optimization problems in the area of software engineering, and it is possible that the results of this paper
will not apply to those models.   Future research should explore
more models to check the validity of our results.

\section{Related Work}\label{sect:rel_work}
{
We introduced a baseline method to solve the SBSE problems through sampling. Many researchers tried to solve the tricky or computationally expensive problems through sampling and other strategies in other domains. \resptof{2a}

For example, sampling has been successively applied to  the noisy real-word optimization problems by
Cantu-Paz~\cite{cantu2004adaptive}. They introduced an adaptive sampling policy which they
test on a   100-bit onemax function. While Cantu-Paz demonstrated that the adaptive sampling could find better solutions,
from our perspective, the drawback with that work is that it requires far more computation time.

% XXX \footnote{XXXX}: so all these are better than us>
Shahrzad et al.~\cite{shahrzad2016estimating} analyzed the advantage of age-layering in an evolutionary algorithm as well. In aged-layered evolutionary algorithms, a small sample of candidates are evaluated first; and if they seem promising, they are evaluated with more samples.
The age-layering method effectively reduces the fitness evaluations and speedups the evolution process. However, 
at least for the aged-layered algorithm reported by Shahrzad et al.~\cite{shahrzad2016estimating}, this approach
still requires millions of model evaluations, Hence, it would not be a candidate for a baseline SBSE method.

% \footnote{so it is better than us?}.

(1+1) EA is another strategy which can reduce the computing intensity of evolution algorithms~\cite{droste2002analysis}. In (1+1) EA, the population size is set to one. The candidate is mutated in some probability and then replaces the former one if better fitness is found. Compared to the common evolution algorithms which population size can up to hundreds or even thousands, the (1+1) EA can significantly reduce the fitness evaluations~\cite{droste2002analysis}. But the
drawback for standard (1+1) EA is that it did not naturally handle models with multi-objectives, or conflicting objectives, which are very common in SBSE.

Another strategy to speed up the evolutionary algorithms is the use of a surrogate model. Ong et al.~\cite{ong2003evolutionary} presented a parallel evolutionary algorithm which leverages surrogate model for solving the computational expensive design problems. 
A surrogate model is a statical model  built to approximate the computationally expensive model. 
They created a surrogate model basing on the radial basis function. The computation of RBF is much cheaper than the original model. But the precise of surrogate model strongly depends on the evaluated candidates. 
To improve the precision of surrogate model for search-based software engineering problems, we have to enlarge the number of model evaluations~\cite{shahrzad2016estimating}.
}

\section{Conclusions and Future Work}
\label{sect:conclusion}
Wolpert et al.~\cite{Wolpert:1997} caution  no optimization
algorithm always works best for all domains.
Hence,   when encountering a new domain, multiple methods should be applied.

When applying multiple methods, it is useful to have a very fast
{\em baseline method} to try first since:
\bi
\item
That offers a useful baseline which can be used
to understand the relative value of other methods.
\item
If this
initial baseline method achieved adequate results, the search through the
other methods can stop sooner.
\ei
For these reasons, many researchers in the field of SE~\cite{Whigham:2015,mittas13,shepperd12z,Kitchenham2007} and elsewhere~\cite{cohen95,holte93} note 
that any field that conducts empirical experiments with algorithms can utilize baseline methods.
Such baselines allow for early feedback about whether or not the optimization is correctly integrated into the model.
They can also be used as {\em scouts} that run ahead of more expensive processes to report the complexity of up-coming tasks. 
For example, Shepperd and Macdonnel~\cite{shepperd12z}
argue  convincingly  that  measurements  are  best  viewed  as  ratios compared  to  measurements  taken  from  some  minimal  baseline
method.

In this paper, we introduced a baseline method, {\sc Sway}, to explore optimization problems
in the context of search-based software engineering problems. \sway can find promising individuals among a large set of candidates using a very small number of model evaluations. Since the number of required model evaluations is much less than that of common evolutionary algorithms, \sway terminates very early.
 \sway would be especially useful when the model evaluation is computation expensive (i.e., very slow). 
 
\sway satisfies all the criteria of a baseline method, introduced in \tion{baseline}; e.g., simple to code,
applicable to a wide range of models.
This paper tested   \sway via numerous scenarios within three SE models.
These models differed in their type of decisions as well as the size of their decision space. Results showed the quality of outputs from \sway were comparable to the state-of-the-art evolutionary algorithms for those specific problems.
\sway is also very fast. Among 15 cases studied in this paper, \sway only requires less than 5\% (in median) of the runtime of the standard evolutionary algorithms, but in majority case across all performance measures, \sway 
performs the same or better as the state-of-the-art.

Extending this work in several ways is possible. In this paper,
we have explored three SE models from the areas of effort estimation, project management as well as requirement engineering.
There are many other domains in SE that might benefit from this approach such
as   testing, debugging and cloud environment configuration, etc. \respto{2-4}{{For example, Wang et al.~\cite{wang2017optimal}
introduced a regression test selection for service-oriented workflow applications. 
We also have further research in configuring the workflows into cloud environment basing on \sway~\cite{chen2017riot}.
In our future work, we will explore the sampling techniques in service-oriented workflow applications as well as other SE models.}}

Second, we introduced the \sway for two types of decision space -- continuous decision spaces (\tion{split_con}) and binary decision spaces (\tion{split_bin}). In the future, we will explore more types of problems, for example, the graph-based models like software modularization~\cite{harman2005empirical}.

Third, the parameters of \sway, such as  ``enough'' in Algorithm \ref{alg:split_con}, were set manually in this paper. Discussion of relations between these parameters and final results is left for future work.

Finally, in \fig{boxplot} we can find that the {\sc GroundTruth} method is almost always comparable to the MOEA methods. But the \sway is beat by the MOEA in some cases. This is the tradeoff of \sway's fast termination. Can we increase the number of model evaluation in some strategy to improve the quality? This is worth exploring.

\bibliographystyle{plain}
% \bibliography{main}

% \balance
\vspace{2mm}

\begin{IEEEbiography}[{\includegraphics[width=1.0in,height=1.1in,clip]{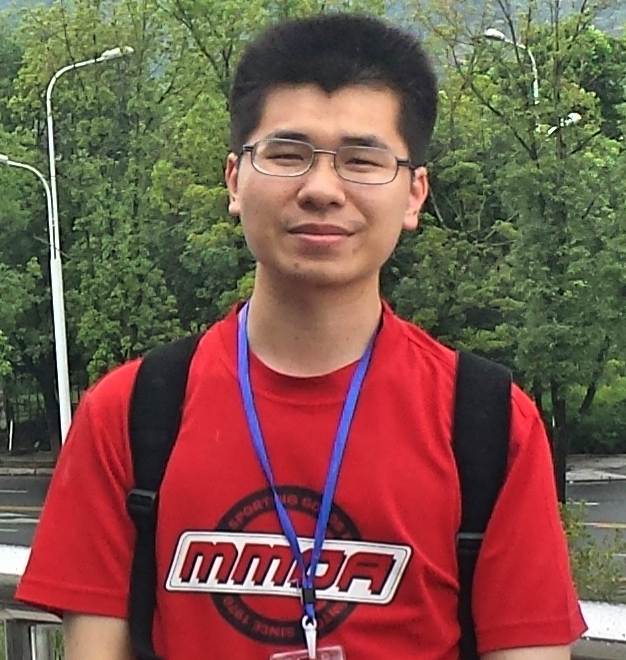}}]
{Jianfeng Chen}  is a computer science PhD candidate at North Carolina State University. He received his bachelor degree in Shandong University, China. His research relates to automated  software  engineering,  artificial  intelligence and  multi-objectives  optimization (with a particular focus   on utilizing  stochastic  sampling  techniques  to  speed  up search-based software engineering solvers).  He also explores methods relating to  fuzz testing, regression testing and  cloud  environment  configurations.  He  has  published papers in IEEE Transactions on Software Engineering, Information Software Technology, International Symposium  on  Search  Based  Software  Engineering. For more, see his homepage \url{http://jianfeng.us}.
\end{IEEEbiography}

\begin{IEEEbiography}[{\includegraphics[width=1.0in,height=1.1in,clip]{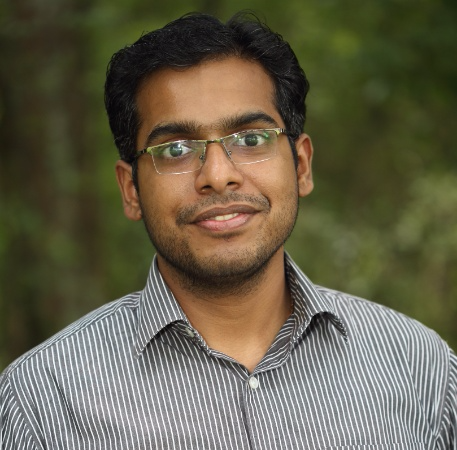}}]
{Vivek Nair} is a fifth year Ph.D. student in department of Computer Science at North Carolina State University. He received his bachelor and master degree in West Bengal University of Technology and National Institute of Technology, Durgapur respectively. His primary interest lies in the exploring possibilities of using multi-objective optimization to solve problems in Software Engineering. He is currently working on performance prediction models of highly configurable systems. He received his master degree 
and worked in the mobile industry for a period of 2 years before returning to graduate school. For more information, visit \url{http://vivekaxl.com}.
\end{IEEEbiography}

\begin{IEEEbiography}[{\includegraphics[width=1.0in,height=1.1in,clip]{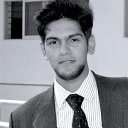}}]
{Rahul Krishna} is a third year Ph.D. student in the department of Computer Science at North Carolina State University. He received his master's degree in Electrical Engineering also at North Carolina State University. His primary interest lies in exploring ways in which aspects of pattern recognition and artificial intelligence can be used to generate actionable analytics for software engineering. He currently works on developing machine learning algorithms that go beyond prediction to generate insights to assist decision making. He also works on transfer learning methodologies that can help transfer models between software projects. For more information, visit \url{http://rkrsn.us}.
\end{IEEEbiography}

\begin{IEEEbiography}[{\includegraphics[width=0.9in, height=1.2in, clip]{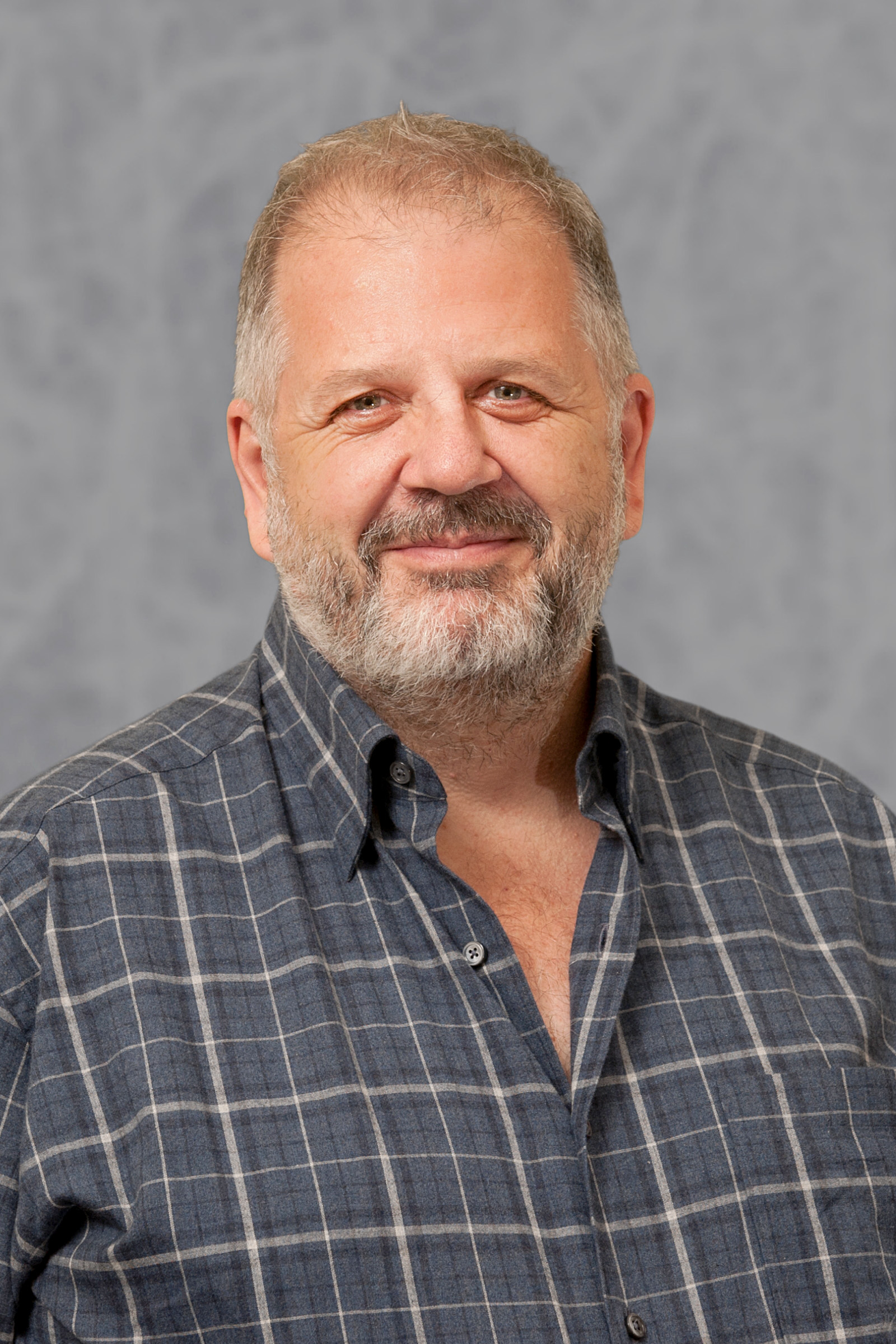}}]
{Tim Menzies} (Ph.D., UNSW, 1995) is a full Professor in CS at North Carolina State University, where he explores SE, data mining, AI, search-based SE, and open access science. He is the author of over 250 referred publications and co-founder of the PROMISE conference series devoted to reproducible experiments in SE (\url{http://tiny.cc/seacraft}).  Dr. Menzies also serves as associated editor of many journals:  IEEE Transactions on Software Engineering (2010 to 2016), ACM Transactions on Software Engineering Methodologies, Empirical Software Engineering, the Automated Software Engineering Journal the Big Data Journal, Information Software Technology, IEEE Software, and the Software Quality Journal. He has served as co-general chair of ICSME'16 and  co-PC chair for ASE'12,  ICSE'15, SSBSE'17. For more, see \url{http://menzies.us}.
\end{IEEEbiography}
\end{document}